%% file: main.tex
\documentclass{article}

\usepackage{arxiv} 
\usepackage[utf8]{inputenc} 
\usepackage[T1]{fontenc}    
\usepackage{hyperref}       
\usepackage{url}            
\usepackage{booktabs}       
\usepackage{amsfonts}       
\usepackage{nicefrac}       
\usepackage{microtype}      

\usepackage{amsmath}        
\usepackage{amssymb}        

\usepackage{subcaption}

\usepackage{graphicx} 

\usepackage{tabularx}
\usepackage{float}

\title{AutoRASOR: Autonomous Rapid Scanning Electron Microscope Operator}

\author{
  \makebox[\textwidth][c]{
    \begin{tabular}{c}
      \textbf{Kevin Zhang}$^{1, 2}$, \textbf{Mohammad Taha}$^1$, \textbf{Rafael Espinosa Casta\~neda}$^{1, 2, 3}$, \textbf{Yutong Liu}$^{1, 4}$, \textbf{Yin Zhu}$^{1, 4}$, \textbf{Lotan Portal}$^1$, \\
      \textbf{Robert A. Mcleod}$^{1, 5}$, \textbf{Vahid Attari}$^{1, 6}$, \textbf{Jane Y. Howe}$^1$, \textbf{Jason Hattrick-Simpers}$^{1, 2, 3, 7, 8}$ \\[0.4cm]
      $^1$Department of Materials Science and Engineering, University of Toronto, Toronto, ON, Canada \\
      $^2$Acceleration Consortium, Toronto, ON, Canada \\
      $^3$Vector Institute, Toronto, ON, Canada \\
      $^4$The Alliance for AI-Accelerated Materials Discovery (A3MD), Toronto, ON, Canada \\
      $^5$Hitachi High-Tech Canada, Inc., Toronto, ON, Canada \\
      $^6$CanmetMATERIALS, Natural Resources Canada, Hamilton, ON, Canada \\
      $^7$Data Science Institute, University of Toronto, Toronto, ON, Canada \\
      $^8$Schwartz Reisman Institute for Technology and Society (SRI), Toronto, ON, Canada \\[0.1cm]
      \texttt{jane.howe@utoronto.ca, jason.hattrick.simpers@utoronto.ca}
    \end{tabular}
  }
}

\begin{document}
\maketitle
\begin{abstract}
\begin{center}
    \includegraphics[width=0.65\textwidth]{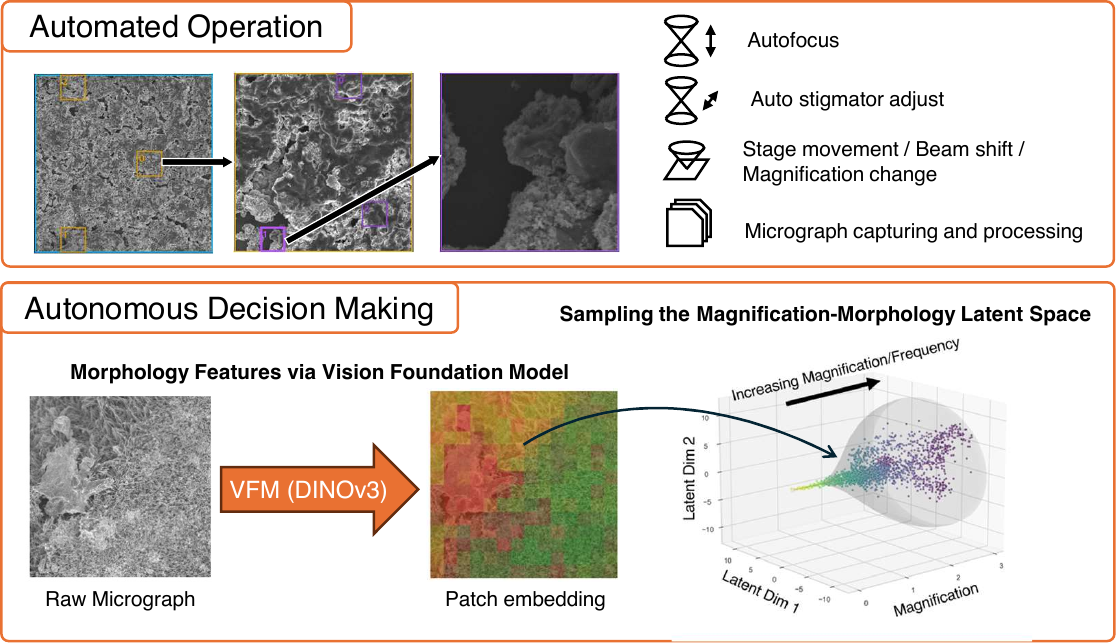}
\end{center}
\vspace{1em}
Scanning Electron Microscopy (SEM) is a foundational technique for characterizing material microstructure, which dictates many fundamental physical and chemical properties. With the rise of Self-Driving Labs (SDLs), samples are now synthesized in large batches, demanding equally high-throughput, autonomous characterization. Existing automated electron microscopy pipelines are task-specific: they detect pre-defined features, optimize known properties, or require prior knowledge of the sample, restricting each pipeline to the material system it was built for. We introduce AutoRASOR, a task-agnostic autonomous SEM pipeline that captures the multi-scale morphology of an unknown sample without domain-specific pre-training, fine-tuning, or human prompting. AutoRASOR embeds micrographs in real time with a vision foundation model (VFM), DINOv3, and selects regions of interest (ROIs) through two complementary policies: Latent Farthest Point Sampling (LFPS) and active learning on morphological ambiguity, which is the conditional variance of unresolved fine-scale features given lower magnification appearance. 
Active learning on ambiguity consistently captures more diverse and rare morphologies than random ROI selection, while LFPS reliably recovers the specimen's morphological distribution within a limited capture budget, tested both on real SEM micrographs and synthetic phase-field images. By shifting autonomous characterization from pre-defined, task-specific targeting of features to morphological survey, AutoRASOR generates information-rich multi-scale datasets for diverse downstream tasks. Without the need for domain-specific pre-training or prior material assumptions, this framework can be deployed out-of-the-box to characterize novel materials produced by SDLs.
\end{abstract}


\section{Introduction}
Microstructure governs many material properties and performance characteristics, making its accurate and representative characterization essential for establishing processing-structure-property-performance relationships. Scanning electron microscopy (SEM) is a foundational characterization technique capable of capturing morphological and compositional information across a large range of magnifications~\cite{goldstein_scanning_2018, mehta_interactions_2012}. However, outside of specialized commercial workflows, conventional SEM characterization remains largely manual, requiring expert microscopists to tune imaging parameters, identify regions of interest (ROIs), and acquire representative micrographs. This dependence on manual operation limits characterization throughput, a challenge that becomes particularly important as materials synthesis itself becomes increasingly automated and high-throughput. 

Self-Driving Laboratories (SDLs) exemplify this shift, using automated experimental design and synthesis to substantially increase the rate at which new material samples are generated~\cite{tom_self-driving_2024, warren_overcoming_2026}. As synthesis throughput increases, characterization can become a bottleneck in the closed-loop materials discovery workflow. Maintaining the benefits of automated experimentation therefore requires characterization methods that can operate with comparable autonomy and throughput. In particular, autonomous SEM workflows capable of selecting informative ROIs and acquiring representative microstructural data with minimal human intervention could enable characterization to scale alongside SDL-driven materials synthesis.

Existing autonomous microscopy frameworks have demonstrated the potential to integrate electron microscopy into autonomous experimental workflows. SDL frameworks like CREST~\cite{zhang_multimodal_2025} and INTERSECT~\cite{engelmann_intersect_2022} have been developed to include autonomous SEM, along with many other pipelines automating related electron microscopy (EM) techniques such as Transmission Electron Microscopy (TEM), Scanning TEM (STEM), and Scanning Probe Microscopy (SPM). Recent autonomous microscopy approaches commonly select regions of interest (ROIs) by detecting or segmenting predefined features \cite{giunto_accurate_2026,wang_sam-em_2025,olszta_automated_2022,clusiau_workflow_2025}, or by using active learning and Bayesian optimization to target specific measurable properties \cite{pratiush_scientific_2025,vatsavai_curiosity_2025}. Commercial automation software similarly provides modular workflows combining functions such as image segmentation, manual ROI selection, autofocus, stage movement, magnification control, and image acquisition~\cite{corporation_automation_nodate}. Although effective for their intended applications, these approaches often rely on prior knowledge of the target features, morphology, property, or acquisition objective. In addition, many existing pipelines operate at a fixed magnification or between two predefined magnifications~\cite{giunto_accurate_2026,wang_sam-em_2025}, with the imaging scale selected based on the expected feature size of the sample. The capabilities and limitations of representative autonomous electron microscopy pipelines are summarized in Table~\ref{tab:em_pipeline_comparison}. There remains a need for an efficient, task-agnostic autonomous SEM pipeline capable of characterizing unfamiliar samples across multiple magnifications without relying on predefined morphological targets or property-based objectives.

\begin{table}[htbp]
\centering
\caption{Comparison of different autonomous electron microscopy pipelines.} 
\label{tab:em_pipeline_comparison}
\newcolumntype{L}[1]{>{\hsize=#1\hsize\raggedright\arraybackslash}X} 
\renewcommand{\arraystretch}{1.3}
\begin{tabularx}{\textwidth}{L{1.0} L{1.1} L{1.1} L{0.8}}
\toprule
\textbf{Pipeline} & \textbf{Prior knowledge required} & \textbf{ROI selection policy} & \textbf{Magnification handling} \\
\midrule
Backscattered Electron (BSE) contrast segmentation for EDS: TESCAN TIMA\cite{noauthor_advanced_nodate}, AutoEMXSp\cite{giunto_accurate_2026}  & Requires compositional contrast on BSE detector  & Zoom into bright regions in BSE image  & Two magnifications \\ 

Particle-segmentation-driven pipelines~\cite{olszta_automated_2022, clusiau_workflow_2025}  & Pre-trained or few-shot ML segmentation model for samples with known particle morphology & Zoom and perform grid capture over segmentation result  & Two magnifications \\ 

Deep Kernel Learning on measured property~\cite{vatsavai_curiosity_2025, pratiush_scientific_2025} & Property-driven, requires and targets measurable physical property and reward & Capture areas that maximize the measured property & Fixed magnification \\ 

Zero-shot VFM detection and segmentation pipelines~\cite{yang_zero-shot_2025, wang_sam-em_2025} & Known material class and target feature, SAM-based detection & Zoom into detected features & Two magnifications \\  

\textbf{AutoRASOR (Ours)} & None, designed for characterization of unknown specimen & Determined on-the-fly, acquisition decisions are driven directly by general-purpose vision embeddings & Multiple discrete magnifications \\ 
\bottomrule
\end{tabularx}
\end{table}

Many of these autonomous EM pipelines utilize Vision Foundation Models (VFMs) for feature extraction. VFMs have emerged as powerful image encoders capable of extracting rich features due to their extensive pre-training on large, diverse datasets. The DINO architecture family is a widely used class of VFMs, with the recent DINOv3 model demonstrating state-of-the-art (SOTA) performance across a wide range of tasks~\cite{simeoni_dinov3_2025, fuster-barcelo_are_2026}. In the microscopy domain, these models are used either directly or via fine-tuning for tasks such as micrograph segmentation, denoising, and super-resolution~\cite{he_unifying_2026, gonzalez-marfil_dinosim_2025}. For image segmentation via VFM, the Segment Anything Model (SAM)~\cite{kirillov_segment_2023} is popular for zero-shot segmentation~\cite{yang_zero-shot_2025, barakati_sam_2025}. Existing frameworks leverage VFMs in two distinct ways. First, they are utilized in post-acquisition analysis for downstream segmentation, clustering, or classification~\cite{yang_zero-shot_2025, barakati_sam_2025, archit_segment_2025}. Second, when deployed live in the acquisition loop, these pipelines rely on segmentation masks or object detection from models like SAM to execute pre-defined, sample-specific tasks, such as adjusting Focused Ion Beam (FIB) parameters~\cite{geurts_llm-driven_2025}, tracking particles in liquid TEM~\cite{wang_sam-em_2025}, and performing grid scans over segmented areas~\cite{olszta_automated_2022, clusiau_workflow_2025}. In both cases, the VFM acts as a feature detector and the acquisition decision is made based on fixed rules built on prior understanding of the sample. There remains a lack of real-time automated microscopy pipelines that directly use the dense VFM embeddings to drive closed-loop acquisition decisions based on the observed morphology of the sample.

To characterize a specimen we aim to map its morphological manifold, defined here as the distribution of observed morphologies across spatial locations and magnifications. The structure of the morphological manifold and the multi-scale sampling problem can be better understood in frequency space. Capturing a micrograph samples the specimen's underlying signal through a bandpass filter whose lower cutoff is set by the field of view (FOV) and whose upper cutoff is set by the spatial Nyquist frequency, and changing magnification shift this sampled frequency band. As magnification increases, the FOV shrinks and the Nyquist limit rises, revealing high-frequency features that are physically unresolvable at lower magnifications. The morphological distribution expands and fragments with magnification, being narrow and unimodal at low-magnification, where most specimens appear macroscopically homogeneous, but can split into distinct clusters of unequal density as magnification rises. As the FOV also shrinks quadratically, more captures are needed to cover the same region size. Therefore, random or grid sampling at high-magnification is inefficient, as it requires more micrographs to recover the true morphological distribution and tends to oversample common morphologies while missing rare ones. Because each higher magnification reveals frequencies absent from the lower magnification, the mapping of morphological features from low to high-magnification is one-to-many. Therefore, low-magnification features could have weak predictive power over high-magnification features. We quantify this nondeterministic mapping as ambiguity, the conditional variance of high-magnification features given a low-magnification neighborhood. 

We developed \textbf{AutoRASOR} (\textbf{Auto}nomous \textbf{RA}pid \textbf{S}EM \textbf{O}perato\textbf{R}) to capture informative, multi-scale micrograph datasets based directly on observed sample morphology without requiring prior knowledge of specimen features. AutoRASOR shifts the objective of autonomous microscopy away from object detection, segmentation, or single property optimization toward comprehensive, task-agnostic sample understanding through diverse sampling of the morphological manifold. By extracting zero-shot semantic feature representations directly from raw micrographs in real time, the pipeline maps microstructural morphology into a latent space without relying on prior material assumptions, fine-tuning, or predefined features, mitigating both human operator sampling bias and task-specific bias. To capture rich morphological representations across magnifications, AutoRASOR selects ROIs using two complementary policies. Latent Farthest Point Sampling (LFPS) spreads captures across the morphology resolved at the current magnification so that common features are not oversampled. Active learning on morphological ambiguity predicts where fine microstructure is poorly resolved by the current view and prioritizes zooming there, avoiding redundant captures of region that the current magnification already resolves.

We introduce first an autonomous, real-time, and closed-loop microscopy framework that directly utilizes VFM embeddings to featurize complex microstructures on-the-fly, enabling unattended acquisition on materials for which no trained model or prior exists. Second, the use of Latent Farthest Point Sampling (LFPS) in the VFM latent space as an ROI selection policy, yielding micrograph datasets that reproduce the specimen's morphological distribution while maintaining diversity in the morphologies captured. Third, a localized morphological ambiguity metric, defined as the conditional variance of high-magnification features given a low-magnification neighborhood and learned by active learning, which allows the identification of areas where zooming in is informative.

\section{Methodology}
\subsection{Microscope Control}
SEM experiments were performed on a Hitachi SU7000 FE-SEM located at the Open Centre for the Characterization of Advanced Materials (OCCAM). SEM automation is achieved via Python scripting using PyAutoGUI to control the GUI and Hitachi's hihi API package for direct instrument control. Custom automated scripts for optical alignment, high-magnification autofocus, high-magnification auto stigmation adjustment (autostig), automatic micrograph capturing and reading, stage control, and beam shift are implemented to run locally on the SEM. Our custom automated SEM allows automatic tuning of focus and stigmation, and acquisition of micrographs at different magnifications and locations across the sample.

\subsection{AutoRASOR Pipeline Overview}
\begin{figure}
    \centering
    \includegraphics[width=0.9\linewidth]{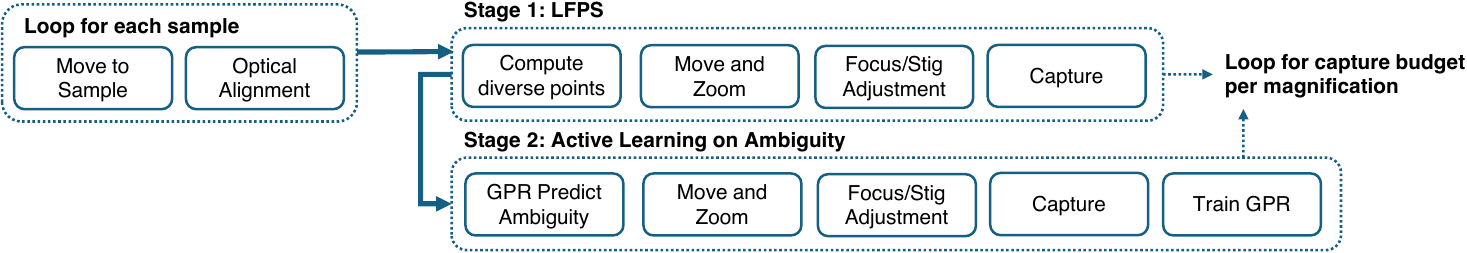}
    \caption{AutoRASOR pipeline for multi-sample, multi-magnification micrograph capturing using LFPS and active learning on ambiguity.}
    \label{fig:autoRASOR_pipeline}
\end{figure}

The overall AutoRASOR pipeline is illustrated in Fig.~\ref{fig:autoRASOR_pipeline}. The user first loads the samples and turns on the electron beam at desired acceleration voltage and spot intensity. From a camera-captured image of the entire specimen, the user then selects the sample location and indicates a micrograph capture budget. The capture budget specifies the number of micrographs requested per sample at low, medium, and high magnification. The initial low magnification can also be set. AutoRASOR then runs unattended, performing optical alignment at every sample, selecting ROIs, and performing autofocus and autostig before capturing every micrograph autonomously. Upon completion, the user turns off the beam and unloads the specimen. The acquired micrographs and associated metadata can then be exported. The AutoRASOR micrograph viewer provides visual inspection of captured micrographs and their spatial position, the relationship between lower and higher magnification micrographs, DINOv3 embeddings, and predicted ambiguity of every single micrograph.

AutoRASOR selects ROIs in two stages at each magnification transition. In stage 1, LFPS selects a set of ROIs that are morphologically diverse at the currently observed magnification, and zooms into each ROI for micrograph capture at the next higher magnification. These data collected in stage 1 provide both an initial mapping between magnifications and warm-up points for active learning on ambiguity. In stage 2, a Gaussian process surrogate trained on these pairs predicts ambiguity across all unobserved regions, and a batched acquisition function selects the next regions to zoom into; each new capture is added to the archive and the surrogate is updated. This two-stage cycle repeats at each magnification level. LFPS and the ambiguity metric are detailed in Sections~\ref{sec:LFPS} and~\ref{sec:ambiguity}.

\subsection{Vision Foundation Model}
We employ DINOv3~\cite{simeoni_dinov3_2025}, a VFM from Meta pretrained on over a billion images, for micrograph featurization. The vits16plus (ViT-S+) variant, the second smallest model in the DINOv3 family, was selected for its processing speed. While massive, domain-specific models require extensive, expensive pre-training on curated custom datasets and significant compute during inference, smaller distilled VFM variants can generate high-quality semantic embeddings close to their non-distilled versions~\cite{simeoni_dinov3_2025}. The model is exported via ONNX runtime\cite{onnx_runtime_developers_onnx_2021} for fast inference on the SEM computer CPU. 

\begin{figure}
    \centering
    \includegraphics[width=0.8\linewidth]{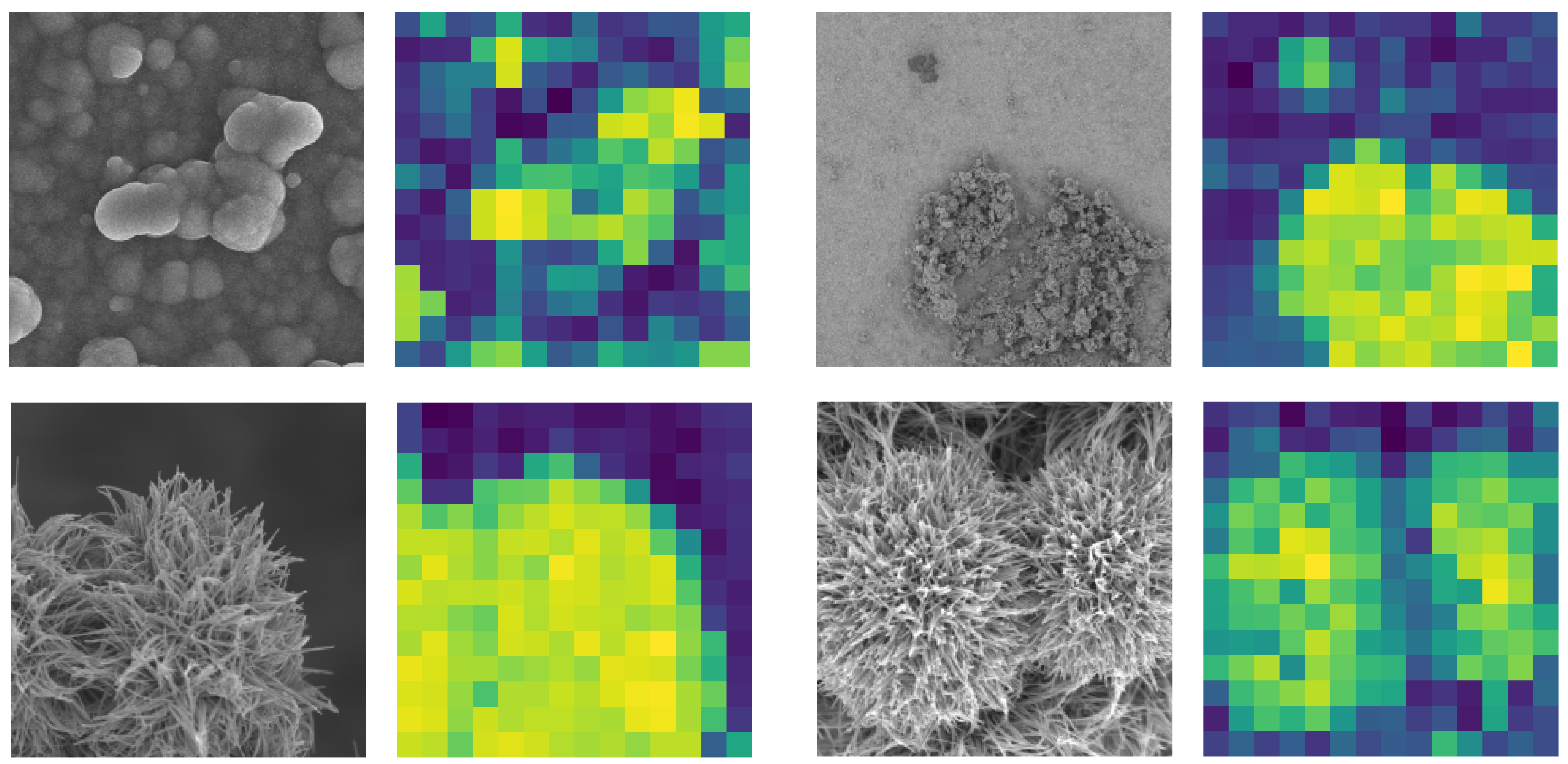}
    \caption{\textbf{How DINOv3 views a micrograph.} Visualization showing the DINOv3 patch similarity for four micrographs. Left: An SEM micrograph down-sampled to 224$\times$224. Right: A heatmap showing the first principal component of each patch. Similar color denotes similar patch embeddings, and therefore similar morphology.}
    \label{fig:DINOv3_PCA_RGB}
\end{figure}

To compute DINOv3 embeddings, micrographs are resized to $224 \times 224$ pixels, a typical resolution used for VFMs and in the training of DINOv3~\cite{simeoni_dinov3_2025}. Passing the downscaled micrograph through DINOv3 generates a tensor of dimension $14 \times 14 \times 384$, yielding a 384-dimensional embedding for each $16 \times 16$ pixel patch. Because DINOv3 is trained such that features reside on a hypersphere~\cite{chang_hyperspherical_2026} in a high-dimensional space, we use cosine similarity to evaluate morphological similarity (latent space distance) between patches. Since DINOv3 encodes positional information~\cite{darcet_vision_2024} and exhibits a higher proportion of positional relative to semantic morphological information in later layers~\cite{ornek_foundpose_2024}, we extract representations from Layer 5, an early layer with minimal positional bias (SI Section~\ref{si:pos_bias}). Fig.~\ref{fig:DINOv3_PCA_RGB} shows the DINOv3 embeddings of a representative micrograph, where each patch is colored by its first principal components. We observe that DINOv3 consistently captures morphological similarities and differences that align with human visual judgment.

\subsection{Latent Farthest Point Sampling (LFPS)} \label{sec:LFPS}
LFPS aims to sample a set of local patches that are the most diverse at the currently observed magnification. Given only the morphology visible in the current micrograph, it targets morphologically distinct local regions. DINOv3 already partitions an input image, the microscope's current FOV, into a $14\times14$ grid of patches, each with a 384-dimensional embedding $z_p$. Each patch is a physical sub-region of the FOV, so any patch can be zoomed into and captured at a higher magnification. The candidate pool $\mathcal{P}$ is the 196 patches of the current micrograph, and the task is to select a budgeted subset whose embeddings are maximally spread out in the DINOv3 latent space.

We build this subset greedily~\cite{eldar_farthest_1997}. Beginning from a randomly chosen patch $\mathcal{S}_1=\{p_1\}$, we repeatedly add the patch that is farthest from everything already selected:
\begin{equation}
p_t = \arg\max_{p\in\mathcal{P}\setminus\mathcal{S}_{t-1}}\;
      \min_{q\in\mathcal{S}_{t-1}} d(z_p,z_q),
\qquad \mathcal{S}_t=\mathcal{S}_{t-1}\cup\{p_t\},
\end{equation}
where $d$ is the cosine distance computed on the full 384-dimensional embeddings without dimensionality reduction. Selection stops when $|\mathcal{S}|$ reaches the capture budget, and each selected patch is zoomed into and captured at high magnification.

LFPS reads only low-magnification embeddings. It requires no high-magnification captures, no surrogate model, and no tuning beyond the capture budget. It can therefore be used to generate an initial understanding of low-to-high-magnification mapping, and capture regions that look diverse at low-magnification.

\begin{figure}
    \centering
    \includegraphics[width=0.8\linewidth]{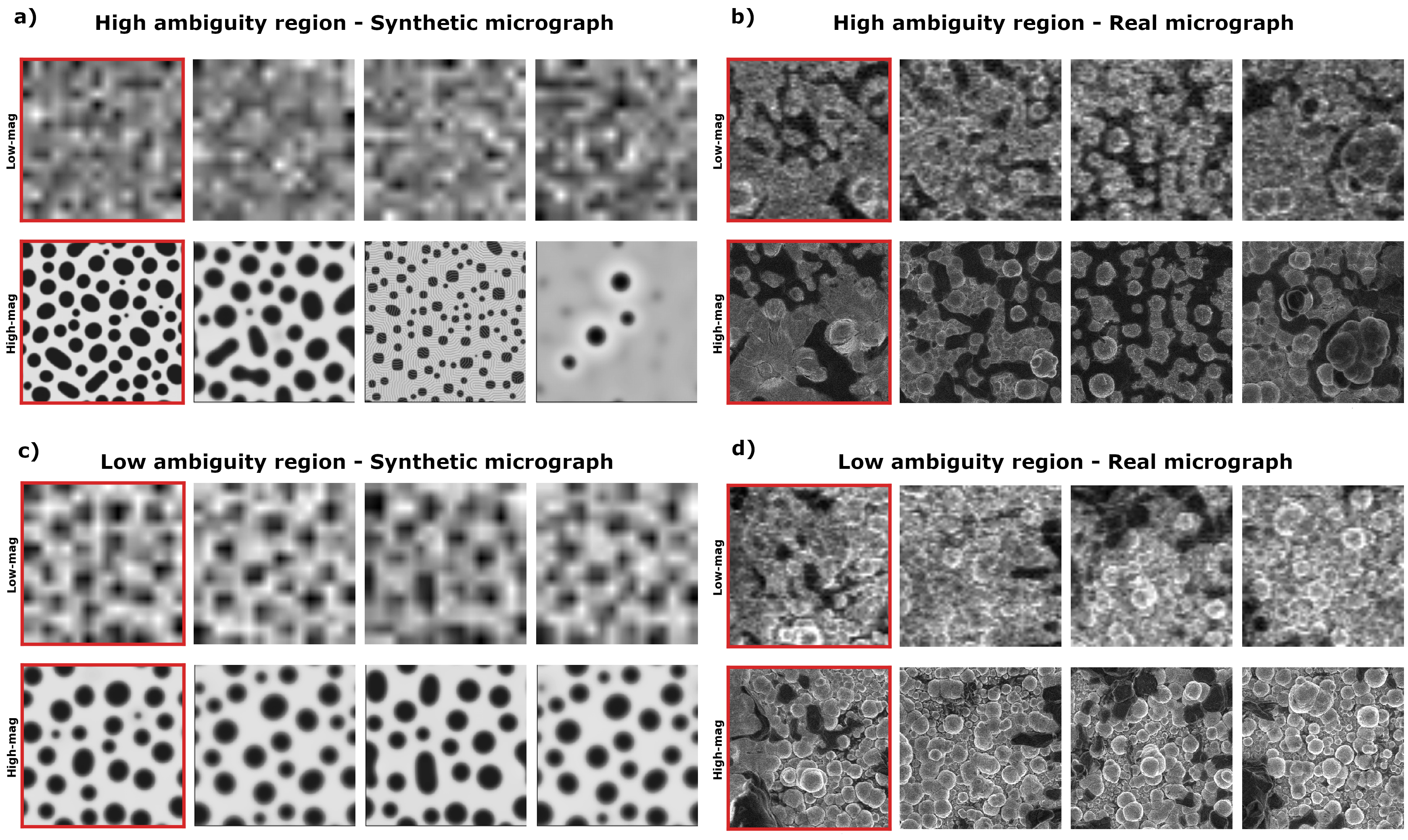}
    \caption{\textbf{High (top) versus low (bottom) ambiguity across synthetic (left) and real (right) micrographs.} Each subplot (a-d) pairs a sequence of visually similar low-magnification patches (top row) with their corresponding high-magnification captures (bottom row). For high-ambiguity patches, their high-magnification captures reveal distinct morphologies despite their low-magnification similarity, indicating the difficulty of predictive low to high-magnification mapping. For example, a) shows four visually similar low-magnification patches resolving into small black particles, very fine black particles on a patterned background, and a few black particles on a gray background, and b) shows that a gray region at low magnification can resolve into a large particle agglomerate, a fine particle agglomerate, or a fused flat surface. In low-ambiguity regions, similar low-magnification patches consistently map to identical, highly predictable high-magnification structures. The high-magnification images in c) and d) exhibit much less diversity than the high-magnifications in a) and b). AutoRASOR targets high ambiguity regions to maximize the discovery of diverse morphologies while avoiding the redundant sampling of the same morphology from low ambiguity regions.}
    \label{fig:high vs low amb}
\end{figure}

\subsection{Ambiguity} \label{sec:ambiguity}
We define ambiguity as the conditional variance of high-magnification embeddings ($Y$) given a low-magnification embedding ($X = x_i$).
Formally, the target population quantity (estimand) at a query patch $i$ is
\begin{equation} \label{eq:amb-estimand}
  \mathcal{A}(x_i) \;=\; \mathbb{E}\!\left[\,\bigl\lVert Y - \mathbb{E}[Y \mid X] \bigr\rVert_2^2 \;\middle|\; X = x_i \right],
\end{equation}
the total conditional variance of $Y$ summed over embedding dimensions.
Intuitively, $\mathcal{A}(x_i)$ measures how much high-magnification morphology detail varies among regions that appear visually similar at low-magnification, as this detail exists beyond the Nyquist frequency of the low-magnification.
 
Estimating \eqref{eq:amb-estimand} is inherently a nonparametric problem.
Because $X$ is continuous, no two distinct regions share exactly the same low-magnification embedding, and re-imaging the same region only resamples imaging noise, not the region-to-region variation that \eqref{eq:amb-estimand} targets.
The spread of $Y$ at a query $x_i$ must instead be estimated by pooling over spatially different regions that have similar coarse appearances.
This is precisely the setting of the classic Nadaraya--Watson (NW) kernel variance estimator~\cite{nadaraya_estimating_1964, watson_smooth_1964, fan_efficient_1998}, which we adapt for active acquisition via $k$-nearest-neighbor ($k$-NN) truncation.
All feature representations are $L_2$-normalized for fast computation of $d_{ij}$ which is the cosine distance in the low-magnification DINOv3 latent space.
For a query patch $i$, we identify its $k$ nearest neighbors in the DINOv3 latent space ($k = 10$), and denote $d_{i,k}$ as the distance of $i$ to the $k$-th neighbor, indicating their low magnification morphological similarity.

Each neighbor $j$ receives a Gaussian kernel weight based on its low-magnification distance to the query $i$:
\begin{equation} \label{eq:weights}
  w_{ij} \;=\; \exp\!\left(-\frac{d_{ij}^{2}}{\ell_i^{2}}\right),
  \qquad
  W_i \;=\; \sum_{j=1}^{k} w_{ij},
\end{equation}
where $W_i$ is the total effective kernel support around $x_i$.

Because feature density on the latent manifold is non-uniform, we employ a per-query adaptive bandwidth $\ell_i$ to prevent kernel collapse in sparse regions:
\begin{equation} \label{eq:bandwidth}
  \ell_i \;=\; \min\!\left(d_{i,k},\, \ell_{\max}\right) + 10^{-6},
  \qquad
  \ell_{\max} \;=\; \operatorname{median}\!\left(\{d_{m,k}\}_{m=1}^{n}\right).
\end{equation}
The global ceiling $\ell_{\max}$ bounds the bandwidth in sparse regions to prevent the $k$-NN search from reaching too far in the latent space, which would include neighbors that are visually dissimilar and inflate the ambiguity estimate. In other words, when the ceiling is active, distant neighbors are exponentially down-weighted rather than treated as if they were similar, so that anomaly does not explode in ambiguity score.

The NW estimate of the expected high-magnification appearance $\mathbb{E}[Y \mid X = x_i]$ is the kernel-weighted average of the neighbors' high-magnification features:
\begin{equation} \label{eq:centroid}
  \bar{y}_i \;=\; \frac{1}{W_i}\sum_{j=1}^{k} w_{ij}\, y_j .
\end{equation}
 
Finally, the localized morphological ambiguity $A_i$ is the weighted variance of the neighbors' high-magnification features about this mean, with $y_j$ being the high magnification pooled embedding of the neighbor $j$:
\begin{equation} \label{eq:ambiguity}
  A_i \;=\; \frac{1}{W_i}\sum_{j=1}^{k} w_{ij}\,\bigl\lVert y_j - \bar{y}_i \bigr\rVert_2^{2}.
\end{equation}

This ambiguity score is high when there are large differences between the high-magnification embeddings (high-magnification micrographs look dissimilar), and the weight is large (they look similar at low-magnification). On the other hand, ambiguity would be low for patches that look the same at low and high magnification, indicating a one-to-one mapping and where predicting high-magnification features from low-magnification is easy, hence no need to repeatedly sample this kind of patch. Fig.~\ref{fig:high vs low amb} shows how patches that appear similar at a coarse scale can resolve into distinct fine-scale morphologies and vice versa.

\subsection{Active Learning on Ambiguity} \label{sec:al}
To explore morphological diversity at the next higher magnification, an active learning pipeline is set up to prioritize sampling of high-ambiguity regions.
By Eq.~\eqref{eq:amb-estimand}, $A_i$ estimates the spread of the conditional distribution $P(Y \mid X=x_i)$. This spread may arise from information unresolved at low-magnification due to the effective frequency cutoff or noise, or from a one-to-many mapping between the coarse- and fine-scale microstructures, where distinct fine-scale features appear similar when viewed at lower magnification. Variance does not distinguish a broad unimodal conditional distribution from a multimodal one; however, both cases motivate additional sampling to characterize the range of possible high-magnification morphologies.
 
Evaluating \eqref{eq:ambiguity} requires the high-magnification capture of the query region and of its low-magnification neighbors.
During a campaign, neighbors are therefore drawn from the set of already-captured patches, and $A_i$ can only be evaluated on that archive.
To direct sampling over the unvisited grid, AutoRASOR needs a predictive model of the ambiguity landscape, which is obtained with a Gaussian process regression (GPR) surrogate.
 
$A$ is modeled as a function of the low-magnification DINOv3 patch embedding $x$ using BoTorch ~\cite{balandat_botorch_2020}, with a constant mean and a Mat\'ern-5/2 covariance kernel with automatic relevance determination (ARD), allowing each latent dimension to be scaled independently. During testing comparable performance was observed using either the full $384$-D DINOv3 embedding, or the $10$-D, $25$-D, or $50$-D PCA-reduced embedding pre-fitted on a large micrograph dataset~\cite{aversa_first_2018}. Therefore, the full $384$-D embedding is used, avoiding dependence and bias from the dataset used to fit the PCA. Further details are discussed in SI section~\ref{si:pca_dim}.

Ambiguity estimates are not equally reliable across the grid.
Neighborhood density varies across the grid, and ambiguity estimates are less reliable early in a campaign when the archive is small, forcing the k-NN search to extend farther in the low-magnification latent space. Such neighbors may be less morphologically similar at low-magnification, potentially increasing the variability and inflating the estimated ambiguity. 
The bandwidth ceiling in \eqref{eq:bandwidth} mitigates this by reducing the weight of very distant low-magnification neighbors, but does not fully remove the underlying problem.
We therefore also treat $A_i$ as a noisy observation of the underlying ambiguity field, using a fixed-noise Gaussian likelihood in BoTorch whose per-patch variance $\sigma_i^2$ is the sampling variability of the estimator itself. Intuitively, if the ambiguity estimate for a query patch changes substantially when a single neighbor is added or removed, that estimate is noisier.
 
This variability is measured by leave-one-neighbor-out resampling: removing each of the $m \le k$ contributing neighbors in turn and recomputing \eqref{eq:ambiguity} yields $A_i^{(-j)}$, whose spread is the jackknife variance~\cite{quenouille_notes_1956, efron_jackknife_1981}:
\begin{equation} \label{eq:loo-var}
  V_i \;=\; \frac{m-1}{m}\sum_{j=1}^{m}\bigl(A_i^{(-j)} - \bar{A}_i^{(-)}\bigr)^{2},
  \qquad
  \bar{A}_i^{(-)} \;=\; \frac{1}{m}\sum_{j=1}^{m} A_i^{(-j)} .
\end{equation}
In the standardized label units used by the GP, with $s_A$ the sample standard deviation of the archive's ambiguity values, the observation variance is
\begin{equation} \label{eq:noise-loo}
  \sigma_i^{2} \;=\; \min\!\left(\sigma_{0}^{2} + \frac{V_i}{s_A^{2}},\; 4\right),
  \qquad \sigma_{0}^{2} = 0.01 ,
\end{equation}
so an estimate dominated by one or two neighbors is downweighted, while one that is stable under deletion is trusted.
The floor $\sigma_0^2$ keeps the likelihood well conditioned, and the cap of four label variances marks the point at which an observation is effectively ignored. The value of this upper bound has no observable significance on the performance of active learning, SI section~\ref{si:max_var} discusses this in detail.

To target high-ambiguity regions and update the surrogate, we use the quasi-Monte-Carlo batched log noisy expected improvement (qLogNEI)~\cite{ament_unexpected_2025} acquisition function implemented in BoTorch.
Because multiple morphologically distinct regions within each iteration can exhibit high ambiguity, a batch policy can facilitate exploration of multiple regions within each iteration. Capturing multiple regions per iteration also enables a more efficient acquisition sequence, in which nearby patches can be imaged consecutively, reducing large travel and changes in focus and stigmation.

\subsection{Synthetic Benchmark Dataset for Method Validation}
To develop and benchmark prior to deployment on the SEM, we constructed a synthetic dual-magnification micrograph generator. The purpose of the synthetic micrograph generator is to provide a low-magnification micrograph in which every local region can be queried to reveal a ground-truth high-magnification morphology. Ground-truth microstructures are sourced from the Open Phase-Field Microstructure Dataset (OPMD) version 1.0~\cite{attari_2024_16756795,attari_opmd_2023}. Unlike purely procedural or generative micrograph datasets, OPMD provides microstructures from physics-based phase-field simulations, providing morphologies that are physically plausible for real-world dual-phase materials.

We treat each phase-field image as a high-magnification micrograph, with a resolution of 224$\times$224 pixels. These images are then arranged in a grid to simulate the low-magnification view. Using a resolution of 224 $\times$ 224 pixels with DINOv3 patch size being 16 $\times$ 16 pixels, the low-magnification micrograph is assembled as a 14$\times$14 grid such that each high-magnification image maps to exactly one low-magnification patch. The effective magnification ratio between the two views is $14\times$.

To create spatially smooth variations in morphology, representative of spatial heterogeneity in real material microstructures, images with similar simulation parameters are preferentially placed near one another. We generate two spatially continuous 2D Gaussian noise maps corresponding to two normalized parameters from the phase-field simulation: area fraction ($A_f$) and phase-field interface mobility ($M$). These two noise maps control the arrangement of the patches on the grid. For each grid position $(i, j)$, the algorithm selects a phase-field image $x$ from the OPMD dataset that minimizes the Euclidean distance in the normalized parameter space:

$$\arg\min_{x} \left( (A_{f,x} - A_{f,i,j})^2 + (M_x - M_{i,j})^2 \right)$$

\begin{figure}
    \centering
    \includegraphics[width=0.8\linewidth]{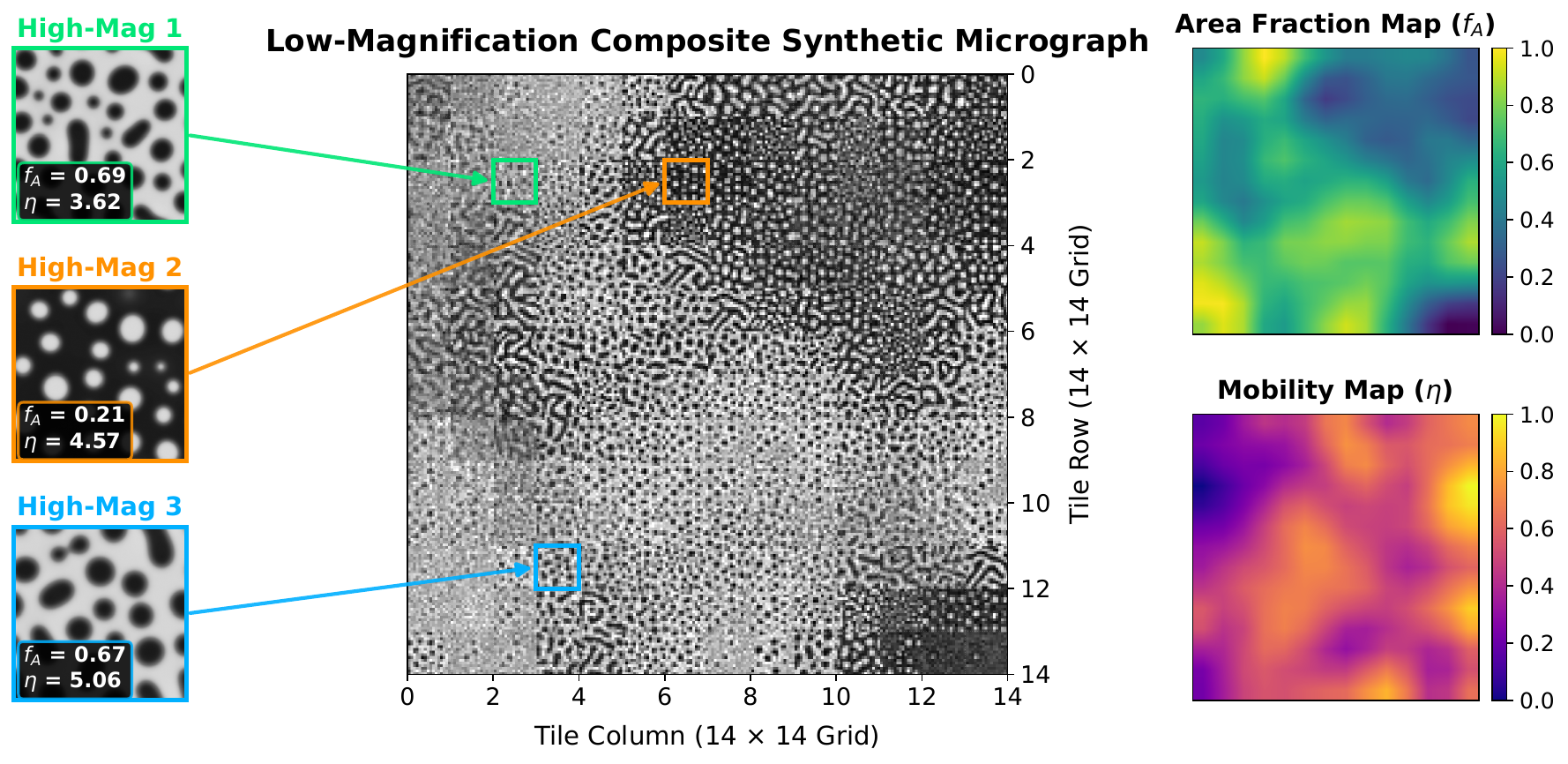}
    \caption{An example of the degraded low-magnification composite micrograph ($14 \times 14$ grid) synthesized by tiling phase-field simulations according to two continuous spatial parameter fields: (Top Right) Area Fraction ($f_A$) and (Bottom Right) Mobility ($\eta$). Simulated SEM degradation (blur, Poisson shot noise, and Gaussian noise) is applied across the image. (Left) High-resolution ground-truth microstructure (High-Mag 1–3) highlighting the morphological variations in this example.}
    \label{fig:synth_example}
\end{figure}

Once the low-magnification composite image is assembled, it is augmented to approximate the signal-to-noise characteristics and imaging artifacts encountered during SEM operation. These augmentations include Gaussian blur, Poisson shot noise, and Gaussian noise. Finally, another spatial noise map is generated to control the local blur intensity across the low-magnification image, simulating defocus associated with variations in specimen height. Fig.~\ref{fig:synth_example} shows one example of the synthetic micrograph dataset generated by this pipeline, and different datasets can be generated by varying the random seed.

The predefined correspondence between each low-magnification view/patch and its original phase-field image provides a ground-truth oracle for evaluating the active-learning campaign. When the acquisition function queries a specific low-magnification patch, the oracle simulates the physical "zooming-in" process by retrieving the corresponding undegraded 224$\times$224 pixel phase-field image, which represents the ground-truth morphology at high-magnification.

\subsection{Validation Testing} \label{sec:validation}
We validate the AutoRASOR ROI selection policies through three complementary evaluations. First, the active learning loop is benchmarked on the synthetic dataset, where ground-truth ambiguity is available to justify the usage of a batched active learning acquisition function. Second, the sampling performance of LFPS and active learning on ambiguity is evaluated on real SEM data against exhaustively captured grid montage datasets. Finally, similar to the second test, LFPS and active learning on ambiguity are tested on the synthetic benchmark for a DINOv3-embedding-independent evaluation in the physical parameter space of the phase-field simulations.

Because the synthetic dataset can return the ground-truth high-magnification image for every region, the true ambiguity of every region can be computed from the complete set of low and high-magnification pairs, allowing direct measurement of how well the active learning loop predicts and samples high ambiguity regions. Each run is initialized with a synthetic low-magnification composite providing 196 queryable regions, and proceeds until 100\% of regions have been sampled, with the surrogate evaluated at 5\% sampling intervals. Four acquisition policies are compared over 10 runs, and within each run the random seed is fixed: random acquisition, LFPS, expected improvement (EI, $q=1$), and batched log noisy expected improvement (qLogNEI, $q=2$), with two additional acquisition policies (qUCB and UCB) evaluated in SI section~\ref{si:al}. Three metrics are reported. Top-20 recall is the overlap between the surrogate's predicted top-20 highest-ambiguity regions and the ground-truth top-20, and measures prediction quality. Top-20 discovery rate is the fraction of ground-truth top-20 regions actually sampled, and measures sampling efficiency. Best ambiguity discovered is the maximum ground-truth ambiguity present in the sampled archive, measuring the efficiency at finding the highest ambiguity patch.
 
To evaluate sampling efficiency on real specimens with real SEM, all high-magnification patches within a set of low-magnification micrographs were captured, producing ground-truth grid datasets against which ROI selection policies can be evaluated. The high-magnification was set to $14\times$ the low-magnification, matching the $14\times14$ DINOv3 patch grid so that each low-magnification patch maps to one high-magnification capture, yielding 196 high-magnification micrographs per region. The complete set of DINOv3 embeddings computed from those 196 micrographs serves as the exhaustively sampled reference  morphological distribution of the observed area, representing every microstructure that could be captured by zooming into this region. Six different regions imaged at various magnifications were captured from two oxygen evolution reaction (OER) catalyst samples: cobalt oxide dropcasted onto Fluorine-doped Tin Oxide (FTO) glass, and Ru-doped manganese oxide on a Titanium grid. The two samples differ in chemistry, substrate, and preparation route, and therefore in morphology and image characteristics.

With these grid datasets captured, the different acquisition policies can then be compared over many random seeds without additional beam time. At each iteration a policy selects a batch of $q=2$ patches and receives their captured high-magnification micrographs. Four policies are compared: random sampling, LFPS, active learning on ambiguity with an LFPS warmup, and active learning without warmup where initial points are selected at random. Active learning with warmup reflects AutoRASOR's deployed workflow, executing LFPS for the first four iterations (eight patches) to construct an initial training dataset for the Gaussian process surrogate before initiating ambiguity-driven acquisition, whereas the cold-started variant selects the initial eight patches randomly instead.
 
We evaluated the diversity of the micrographs captured by each policy on the real SEM grid dataset using three metrics. The variance of the acquired patch embeddings is measured using cosine similarity in the native 384-dimensional DINOv3 space, and is high when the captured dataset spans diverse morphologies. Sliced Wasserstein distance (SWD) measures the optimal transport distance between the sampled embeddings and the full exhaustively sampled reference dataset, and is low when the captured dataset is statistically representative of the region. Local ground-truth sparsity measures the average cosine distance between each sampled patch embedding and its $k$ nearest neighbors within the full dataset ($k = 1\%$ of the dataset size), and is high when rare, isolated morphologies are captured. Because acquiring more than roughly 20\% of a region at high-magnification is time consuming and rarely done in practice, evaluation focuses on early iterations. Each policy is run with 10 random seeds in each of the six regions. Since the six grid datasets contain different morphology, absolute metric values are not comparable between regions, therefore, a rank-based relative score is computed at each iteration within each region. The relative differences in performance between policies are preserved by mapping the best-performing policy to 1 and the worst to 0, with the final scores averaged across the six regions. For variance and local ground-truth sparsity, a higher value is better, therefore the best-performing policy is the one with the highest value. For SWD, a lower value means closer to the reference distribution, so the best-performing policy is the one with the lowest value of SWD. This preserves the ordering of policies while allowing a fair averaging across regions.

A potential concern with the preceding evaluations is circularity, where both ROI selection and the evaluation metrics operate in the DINOv3 embedding space, so a policy that optimizes coverage of that space is favored by design. The synthetic benchmark permits an embedding-independent test, as every patch in the composite low-magnification image originates from the OPMD dataset with known phase-field parameters. Each sampled region is mapped to its area fraction ($A_f$), interface mobility ($M$), gradient energy coefficient (kappa), and characteristic length scale (CLS). Composition sets the phase area fraction, mobility controls feature size and connectivity, kappa controls the sharpness of the phase boundary, and CLS reflects the average size of phases in the microstructure. Parameters are normalized over the full OPMD dataset. The variance, the sliced Wasserstein distance to the full parameter distribution, and the convex hull volume of the sampled parameter vectors are reported, and averaged over 10 random seeds. The rest of the setup is the same as the validation test on the real SEM grid datasets. Convex hull volume replaces the local GT sparsity metric used for real SEM micrograph dataset as the phase-field parameter manifold is continuous without distinct outliers, yielding nearly uniform local sparsity across patches. Conversely, convex hull volume is not used for analysis in DINOv3 latent space due to its high dimensionality, making it not feasible to compute.

\section{Results} \label{sec:result}
\subsection{The Morphological-Magnification Manifold} \label{sec:morph-manifold}
\begin{figure}[!h]
    \centering
    \includegraphics[width=0.95\linewidth]{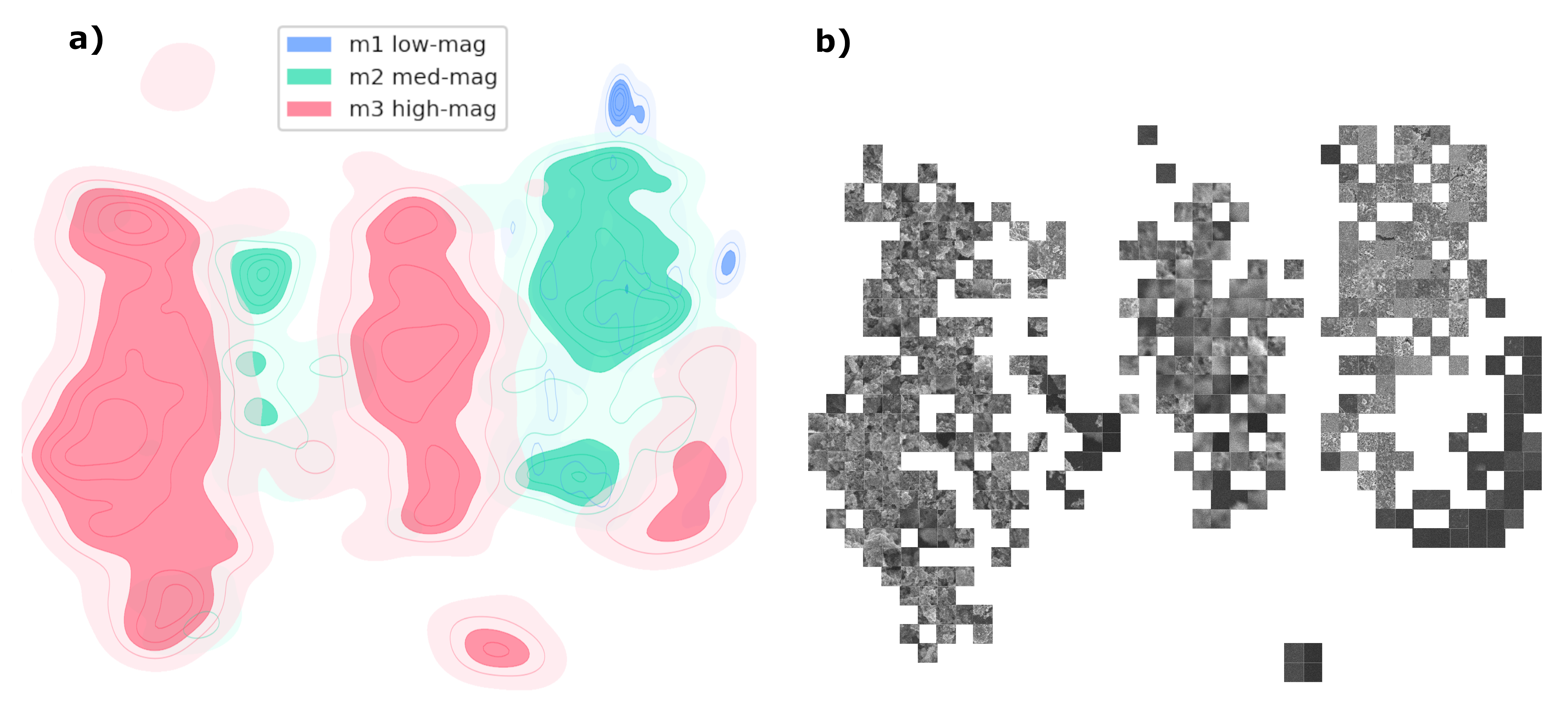}
    \caption{\textbf{How DINOv3 groups similar morphology together across magnification.} a) KDE density plot of the morphological distribution of DINOv3 embeddings from micrographs at three different magnifications. b) Morphology thumbnail plot showing the representative morphology at each UMAP coordinate, the 5x5 patch area around the representative patch is plotted. We observe that similar morphologies are clustered in the DINOv3 latent space.}
    \label{fig:mag-morph manifold}
\end{figure}

A total of 78 micrographs were automatically acquired using LFPS at three magnifications (500$\times$, 3500$\times$, and 24,500$\times$), yielding 15,288 DINOv3 patch embeddings. To visualize the evolution of the morphological representation across magnifications, the embeddings were projected into two dimensions using UMAP, as shown in Fig.~\ref{fig:mag-morph manifold}a. Representative microstructures sampled across the projected space are shown in Fig.~\ref{fig:mag-morph manifold}b. Across all three magnifications, visually similar morphologies occupy neighboring regions of the projection, whereas morphologically distinct features are separated. This organization emerges without material-specific fine-tuning of DINOv3, demonstrating that the pretrained embeddings capture meaningful morphological differences across the investigated magnification range.

As the increase in magnification raises the Nyquist limit and reveals spatial frequencies physically unavailable at lower magnification, the morphological distribution will broaden and separate. Fig.~\ref{fig:mag-morph manifold}a shows this directly. At 500$\times$ the distribution forms two small, compact clusters, as most samples tend to look homogeneous macroscopically. At 3500$\times$ there is one large cluster surrounded by many smaller ones. At 24500$\times$ the manifold becomes a few large connected clusters with varying density. 

\subsection{Active Learning Benchmark}
\begin{figure}[!h]
    \centering
    \includegraphics[width=1\linewidth]{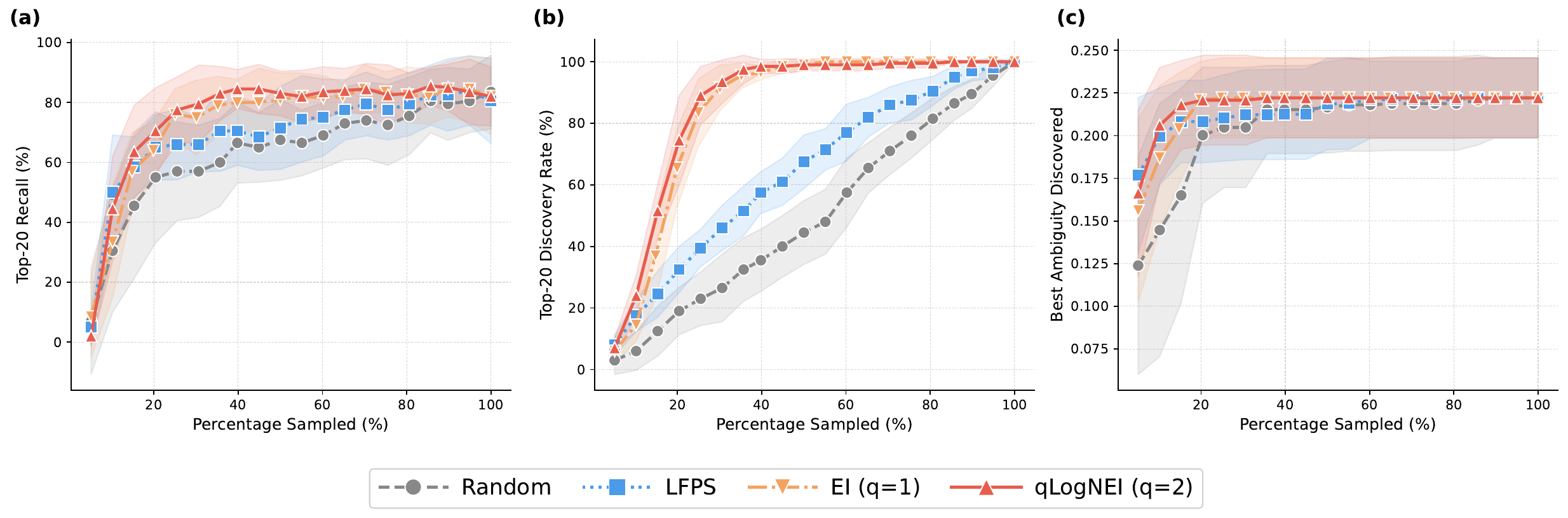}
    \caption{a) The top-20 recall, b) top-20 discovery rate, and c) best ambiguity score discovered so far, averaged over 10 different random seeds. qLogNEI predicts high ambiguity regions better (higher recall) and samples more high ambiguity regions (higher discovery rate) during its active learning campaign than random, LFPS, and its non-batched counterpart.}
    \label{fig:al_synth_result}
\end{figure}

We first evaluate the active learning loop on the synthetic benchmark, following the protocol in Section~\ref{sec:validation}. The performance of the four acquisition policies is shown in Fig.~\ref{fig:al_synth_result}, and the three additional acquisition policies reported in SI section~\ref{si:al} show the same findings. qLogNEI performs comparably to, if not better than, its non-batched counterpart EI at every percentage sampled, showing that batched acquisition does not cost surrogate performance. Both Bayesian acquisition policies predict high ambiguity regions more accurately (higher top-20 recall) and sample them more efficiently (higher top-20 discovery rate) than random and LFPS acquisition. LFPS also outperforms random sampling without requiring high-magnification captures, making it suitable as the warmup stage in our deployed pipeline.

\subsection{Validation on Real Microscopy Data} \label{sec:real-validation}
\begin{figure}[!h]
    \centering
    \includegraphics[width=1\linewidth]{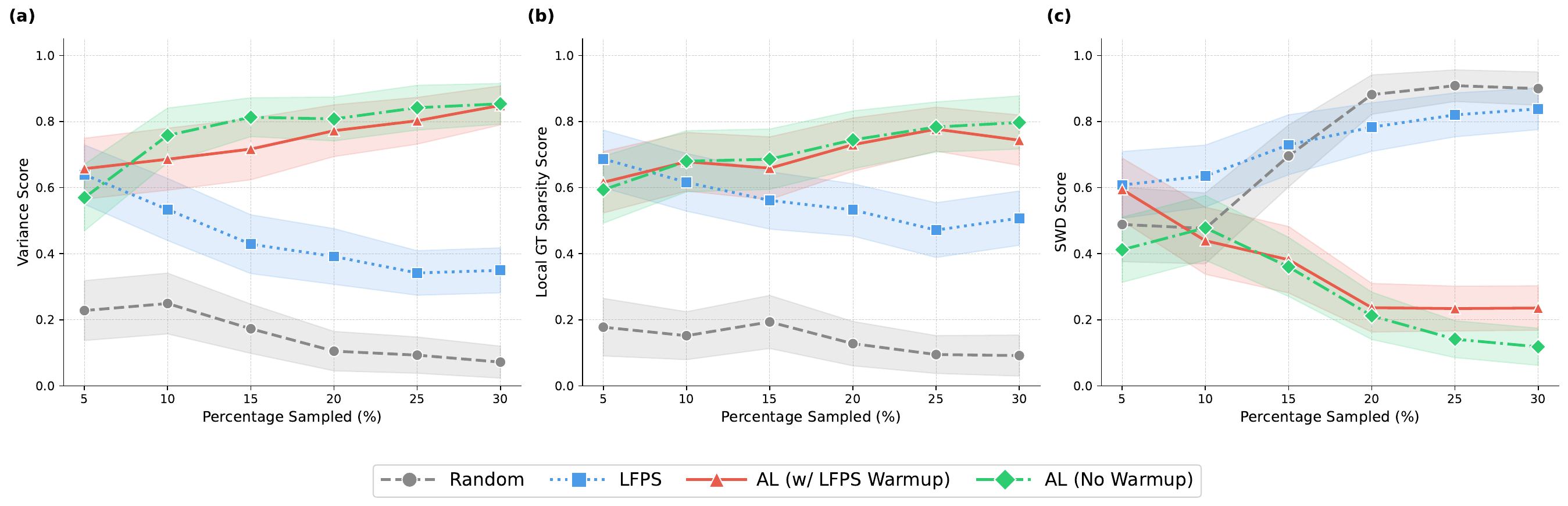}
    \caption{\textbf{Active learning on ambiguity consistently captures more diverse and rare micrographs than random and LFPS across different regions.} Relative scores for a) Variance, b) Local GT Sparsity, and c)SWD are aggregated across 6 distinct regions (each evaluated over 10 random seeds). Active learning policies, both with and without warmup, achieve significantly higher variance and local GT sparsity than LFPS and random baselines, indicating preferential sampling of diverse and rare high-magnification morphologies. Its low SWD score is a consequence of this diversity sampling. LFPS also achieved higher variance and local GT sparsity than random at every iteration, while still maintaining a similar SWD, indicating increased morphological diversity while largely preserving the reference distribution. Shaded regions represent 95\% CI.}
    \label{fig:6_grid_validation}
\end{figure}

We benchmark AutoRASOR's sampling policies on the six real SEM grid datasets described in Section~\ref{sec:validation}, with relative scores reported in Fig.~\ref{fig:6_grid_validation} and results for each individual region reported in SI section~\ref{si:all_grid}. Active learning on ambiguity and LFPS outperform random sampling on the diversity-oriented metrics, but exhibit different behavior with respect to distributional representativeness. Active learning prioritizes regions that reveal diverse and rare morphologies at high magnification, resulting in higher variance and local GT sparsity, with both the warmup and cold-start variants outperforming the baselines on these metrics. However, active learning exhibits lower SWD scores than random sampling and LFPS, indicating lower distributional representativeness. This is consistent with its objective of preferentially sampling distinct and rare morphologies rather than reproducing their prevalence in the reference morphological distribution. In contrast, LFPS exhibits higher variance and local GT sparsity than random sampling while maintaining comparable SWD scores, indicating that it samples more diverse and less common morphologies while largely preserving the reference morphological distribution.

\subsection{Embedding-Independent Validation in Phase-Field Parameter Space}
\begin{figure}[!h]
    \centering
    \includegraphics[width=1\linewidth]{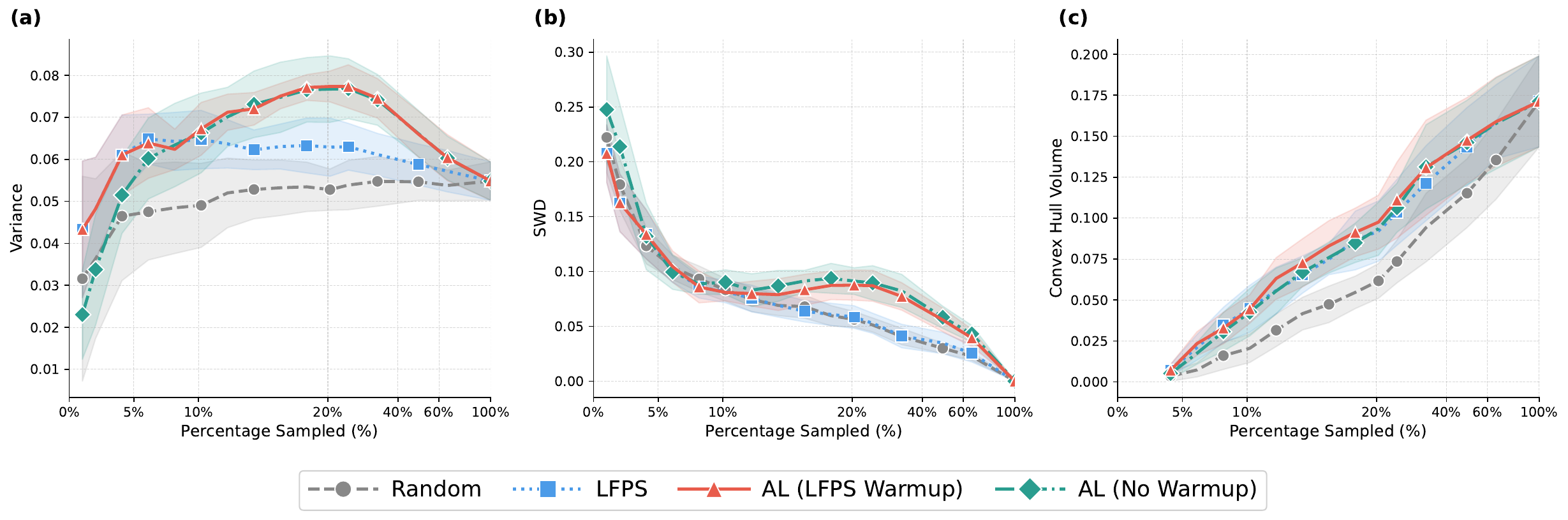}
    \caption{\textbf{AutoRASOR samples phase-field images with diverse and representative physical parameters.} a) Variance, b) SWD, and c) Convex Hull Volume of the normalized composition-mobility-kappa-CLS parameter space on the synthetic benchmark. Active Learning on ambiguity achieves the highest variance and convex hull volume at every budget above 2\% despite selecting solely on DINOv3 embeddings. LFPS achieves higher variance and convex hull volume than random and maintain a similar SWD to the ground-truth reference parameter distribution. LFPS is therefore a strong choice for diverse sampling when statistical representativeness is also desired. The mean over 10 random seeds is plotted and shaded regions show standard deviation. Symlog scaling is used on the x-axis to show performance in earlier iterations clearly.}
    \label{fig:physical_valid}
\end{figure}

To rule out circularity between ROI selection and evaluation, we evaluate the sampling policies directly in the phase-field parameter space of the synthetic benchmark, following the protocol described in Section~\ref{sec:validation}, with results shown in Fig.~\ref{fig:physical_valid}. These parameters are not provided to AutoRASOR during acquisition and are used only for post-acquisition evaluation. Active learning on ambiguity achieves the highest variance at every sampling budget above 2\% and comparable convex-hull coverage to LFPS, despite acquisition being guided solely by DINOv3 embeddings. This result indicates that ambiguity-based sampling promotes broader coverage of the underlying phase-field parameter space without direct access to these parameters. LFPS achieves SWD comparable to random sampling while attaining higher variance and convex-hull coverage, indicating increased parameter-space diversity while largely preserving the reference parameter distribution. In contrast, active learning exhibits poorer distributional representativeness, consistent with its preferential sampling of rare and distinct morphologies.

\section{Discussion} \label{sec:discussion}
The expansion and fragmentation of the morphological distribution with magnification (Fig.~\ref{fig:mag-morph manifold}) directly affect the efficiency of micrograph acquisition. As the FOV shrinks quadratically with increasing magnification, the distribution splits into clusters of unequal density. This multimodal distribution at higher magnification will not be sampled efficiently by sampling based on spatial coordinates. Grid or random capture at high-magnification will mostly land in the dense central cluster, so the dominant morphology is imaged repeatedly while sparse ones are missed entirely. The cost of an exhaustive grid scan also grows quadratically with magnification, making it infeasible to perform at every region on a sample. LFPS and active learning on ambiguity perform acquisition according to the morphological latent space rather than position on the stage. Across the real and synthetic benchmarks, morphology-guided policies increased diversity-related metrics relative to random sampling, including embedding variance and local GT sparsity for the real SEM datasets and variance and convex-hull coverage in the phase-field parameter space (Figs.~\ref{fig:6_grid_validation} and~\ref{fig:physical_valid}). On the synthetic dataset, for random acquisition to capture the same convex-hull-volume coverage as LFPS or active learning on ambiguity in early iterations, it would require an average of 1.7$\times$ more samples and thus time. These results indicate that efficient multi-magnification characterization can be improved by selecting regions based on observed morphology rather than on spatial coverage alone.

The embedding-independent validation indicates that a general-purpose VFM can provide a useful representation for morphology-guided acquisition without explicit knowledge of specimen-specific characteristics, in this case the underlying phase-field parameters. Active learning on ambiguity operates solely on DINOv3 embeddings, yet promotes broader coverage of the phase-field parameter space, achieving high variance and convex-hull coverage, while LFPS increases parameter-space diversity relative to random sampling and closely preserves the reference distribution (Fig.~\ref{fig:physical_valid}). These results suggest that, for the synthetic microstructures investigated here, morphological differentiation in the DINOv3 latent space can also promote coverage of the underlying phase-field parameter space, despite these parameters not being provided to the acquisition policies. This suggests that morphology-guided acquisition may primarily require the ability to distinguish visually different microstructures rather than explicitly identify or classify them, potentially reducing the need for material-specific model training.


LFPS and active learning on ambiguity estimate different statistics of the morphological distribution and are complementary. LFPS reproduces the ground-truth distribution while avoiding redundancy, as reflected in its low SWD and higher variance than random, and is suited to measure quantities that depend on representativeness, such as phase fractions, particle counts, and size statistics. Active learning on ambiguity targets the spread and tail of the distribution at the cost of representativeness. Regions selected for high predicted ambiguity also exhibited high local GT sparsity (Fig.~\ref{fig:6_grid_validation}), indicating that regions with unpredictable high-magnification morphologies tend to be rare morphologies of the specimen. 

By prioritizing high-ambiguity regions, this policy inherently limits redundant zooming. Regions exhibiting low ambiguity represent deterministic, one-to-one cross-magnification mappings where high-magnification captures yield highly predictable microstructure. Ambiguity-based acquisition allocates the capture budget to deprioritize deterministic regions, halting the zoom process there, and redirects it toward regions where fine-scale features remain unresolved.
Ambiguity is suitable for tasks such as assessing homogeneity, detecting anomaly morphologies, or checking batch consistency of a synthesized material at high magnification. Together, LFPS surveys diverse low-magnification regions, while ambiguity targets the unresolved structures revealed only at higher magnification. Because each high-magnification micrograph is recorded together with the low-magnification region it was captured from, the dataset also preserves the macro-to-microstructure relationship that fixed-magnification acquisition pipelines do not provide.

These findings change how a representative set of micrographs can be obtained. In conventional SEM practice, representative characterization commonly relies on systematic spatial sampling, exhaustive montage acquisition, or expert-selected ROIs, which is fast but undocumented and difficult to reproduce between operators or sessions. AutoRASOR specifies the selection procedure with its policy, and each micrograph is stored with the quantity that selected it, either its latent distance from previously captured morphology or its predicted ambiguity. Data collection therefore becomes more reproducible and less dependent on operator ROI-selection decisions, and is capable of running unattended to maximize instrument utilization.

Espinosa Casta\~neda et al.~\cite{castaneda_building_2026} showed that datasets acquired by diversity sampling during active learning in materials science are more useful across downstream tasks and property predictions than datasets acquired for a single objective. Since LFPS and ambiguity acquisition are decoupled from a single scalar material property or feature, the resulting multi-scale survey remains versatile for downstream analysis that may not be anticipated at the time of characterization. This reusability of the dataset is particularly critical as instrument time is limited, specimens can degrade under storage or beam exposure, and re-imaging is often impractical. 

For SDLs designed on material-specific features, exploring novel material design spaces typically requires custom computer vision models to be built, fine-tuned, or prompted for each new material composition. While this framework still requires microscope parameters such as accelerating voltage and spot intensity to be set for the specimen loaded, the acquisition loop itself is material-agnostic. As selection is based on the embeddings from a general-purpose VFM, no segmentation model, property objective, or target feature is needed. The same setup was run without modification on two catalysts differing in chemistry, substrate, and preparation route (Section~\ref{sec:real-validation}). By driving acquisition based on visual differentiation rather than predefined material-specific features, AutoRASOR reduces the need for sample-specific computer-vision model development or retraining, potentially lowering the software adaptation cost associated with deploying autonomous SEM workflows to new material systems.

The current AutoRASOR pipeline has several limitations. First, the performance of the active learning on ambiguity relies on the accuracy of the ambiguity score, which is computed based on the high-magnification micrograph. Therefore, if the high-magnification micrograph is not representative of the true morphology due to imaging artifacts such as charging or contamination, the ambiguity score may be inaccurate and lead to oversampling of a specific artifact. To mitigate this issue, we can inject prior knowledge by filtering out regions that are likely to be affected by these artifacts, such as voids or regions that increase or decrease in intensity/contrast as magnification changes (charging). Second, DINOv3 is trained on natural images and may not be the optimal representation for micrograph features. Micrograph features are often better represented in frequency space, and magnification is not discrete but rather a continuous parameter. Featurizing micrograph in the frequency space would provide a continuous latent space with respect to actual physical scale. This would allow a more natural and continuous expression of ambiguity and eliminate the discretization in mapping from lower to higher magnification. These two limitations both sacrifice the generalizability of AutoRASOR toward a more domain-specific, informed pipeline. 

Third, our real sample validation covers two OER catalysts on different substrates and preparation routes. While these samples, captured at different magnifications, show diverse morphologies and imaging characteristics, they do not span the full range of specimen types that exist in materials characterization, and generalization beyond this range remains to be tested. Finally, the current active learning pipeline only considers the morphological features of the sample, ignoring the sample's chemical composition and synthesis parameters. Incorporating these additional parameters into the active learning loop can expand this pipeline into a multi-sample morphology exploration tool, connecting morphological similarity and synthesis parameter similarity for more efficient acquisition.

\section{Conclusion}
We introduced AutoRASOR, a material-agnostic autonomous SEM pipeline to address the characterization bottleneck in high-throughput materials discovery. By operating directly on semantic embeddings from a Vision Foundation Model, DINOv3, AutoRASOR efficiently maps the morphological-magnification manifold of unknown samples without requiring predefined targets, sample-specific fine-tuning, or human prompting. We demonstrated effective autonomous multi-scale characterization with two complementary sampling policies. LFPS targets diverse low-magnification regions to build a statistically representative baseline distribution without redundant sampling. Active learning on morphological ambiguity targets the unresolved, fine-scale features revealed only at higher magnifications, capturing morphological diversity and rare morphologies. Both policies captured datasets with higher diversity and coverage than random sampling not only in the morphological latent space of DINOv3, but also the phase-field parameter space, demonstrating morphology-guided acquisition using a VFM can effectively explore an unseen material parameter-space.

AutoRASOR reframes autonomous microscopy from a specialized, task-specific measurement tool into a general-purpose, exploratory survey. By standardizing the capture of information-rich, multi-scale micrograph datasets, AutoRASOR reduces human operator bias and generates representative micrograph datasets for diverse downstream analysis. As an out-of-the-box framework that requires no prior assumptions about material composition or feature size, this pipeline provides a highly generalizable characterization module ready for potential integration into closed-loop SDLs.

\section{Acknowledgement}
We thank the funding from NSERC Discovery Grants and Acceleration Consortium Moonshot Grant for supporting this work. This research was undertaken thanks in part to funding provided to the University of Toronto's Acceleration Consortium from the Canada First Research Excellence Fund (grant number CFREF-2022-00042). We acknowledge the support of the Natural Sciences and Engineering Research Council of Canada (NSERC) Discovery Grant, [funding reference number RGPIN-2025-06490, RGPIN-2023-04843].

The authors would also like to acknowledge and thank the support from the Ontario Graduate Scholarship (OGS), the Open Centre for the Characterization of Advanced Materials (OCCAM)  funded by the Canada Foundation for Innovation, and The Alliance for AI-Accelerated Materials Discovery (A3MD). 

\section{Code and data availability}
Code and data are available at: https://github.com/zzy-kevin/AutoRASOR.

\bibliographystyle{unsrturl}
\bibliography{references.bib}

\input{supplement.tex}

\end{document}

%% file: supplement.tex
\clearpage
\appendix

\begin{center}
    \LARGE \textbf{Supplementary Information}
\end{center}
\vspace{1em}

\setcounter{section}{0}
\setcounter{figure}{0}
\setcounter{table}{0}
\setcounter{equation}{0}

\renewcommand{\thesection}{S\arabic{section}}
\renewcommand{\thefigure}{S\arabic{figure}}
\renewcommand{\thetable}{S\arabic{table}}
\renewcommand{\theequation}{S\arabic{equation}}

\section{Positional Bias in DINOv3} \label{si:pos_bias}
DINOv3 encodes positional information in its embedding. While this could be useful for natural image, it is harmful for micrographs as micrographs are textures and are translational invariant. The same feature in the top right of a micrograph should be have different embedding than the same feature in the bottom left of a micrograph. 

\begin{figure}[H]
    \centering
    \includegraphics[width=1\linewidth]{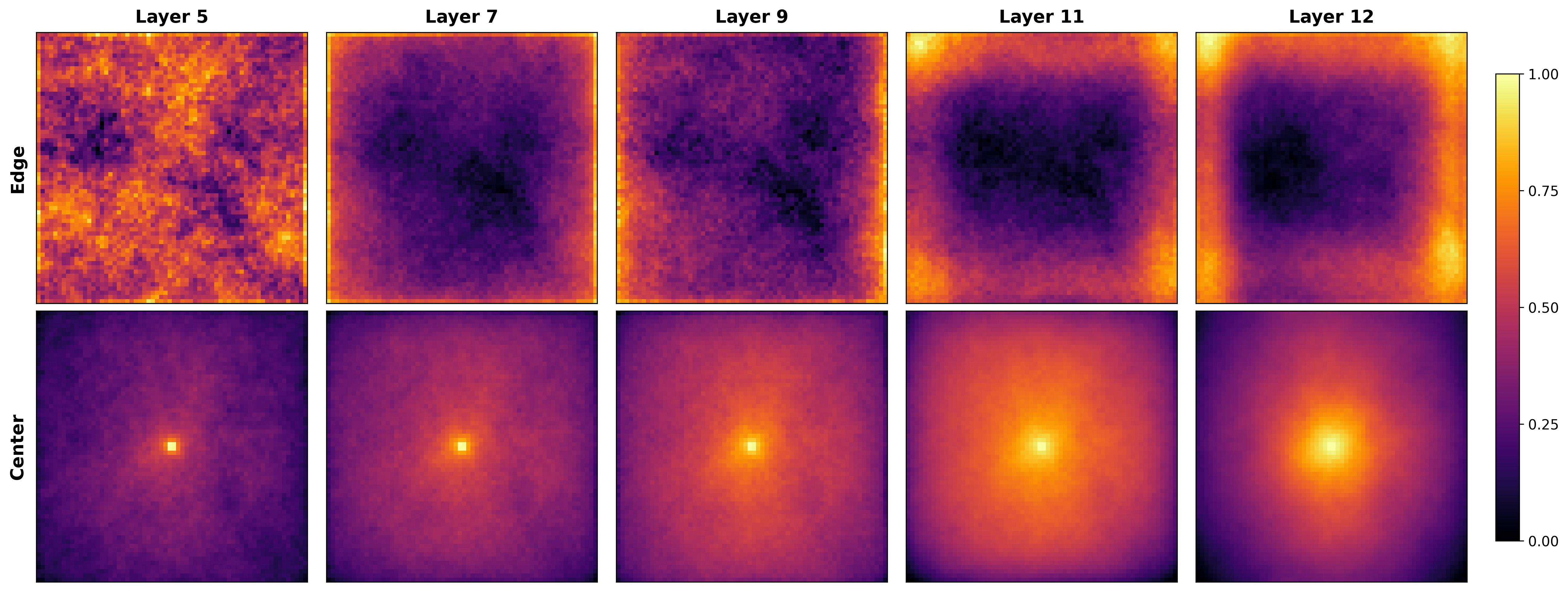}
    \caption{Top: Cosine similarity of each patch to edge patch in the DINOv3 latent space. Bottom: Cosine similarity of each patch to the center patch in the DINOv3 latent space. Results are the average from 45 micrographs. }
    \label{fig:si_pos_bias}
\end{figure}

We observed that positional bias are less severe in earlier layer of DINOv3 for micrographs, matching the conclusion from \cite{ornek_foundpose_2024}. We show this positional bias of layer 5, 7, 9, 11, and 12 (final layer) in Fig. \ref{fig:si_pos_bias}. For the top row, a brighter glow from the edge indicates patches near the edge have similar embedding to the patches at the edge. For the bottom row, a brighter glow in the center indicates patches near the center have a more similar embedding to the center patch. Stronger positional bias, or glow, can be seen in later layer.

\section{Selection of Max Variance in Heteroscedastic Observation Noise} \label{si:max_var}
Active learning on ambiguity uses the jackknife noise estimator, in which the per-observation variance is capped at $\sigma^2_{\max}=4$. Because the GP labels are standardized, this bound is expressed in units of the archive's ambiguity variance, being four times the spread of the ambiguity values themselves. 

\begin{figure}[H]
    \centering
    \includegraphics[width=1\linewidth]{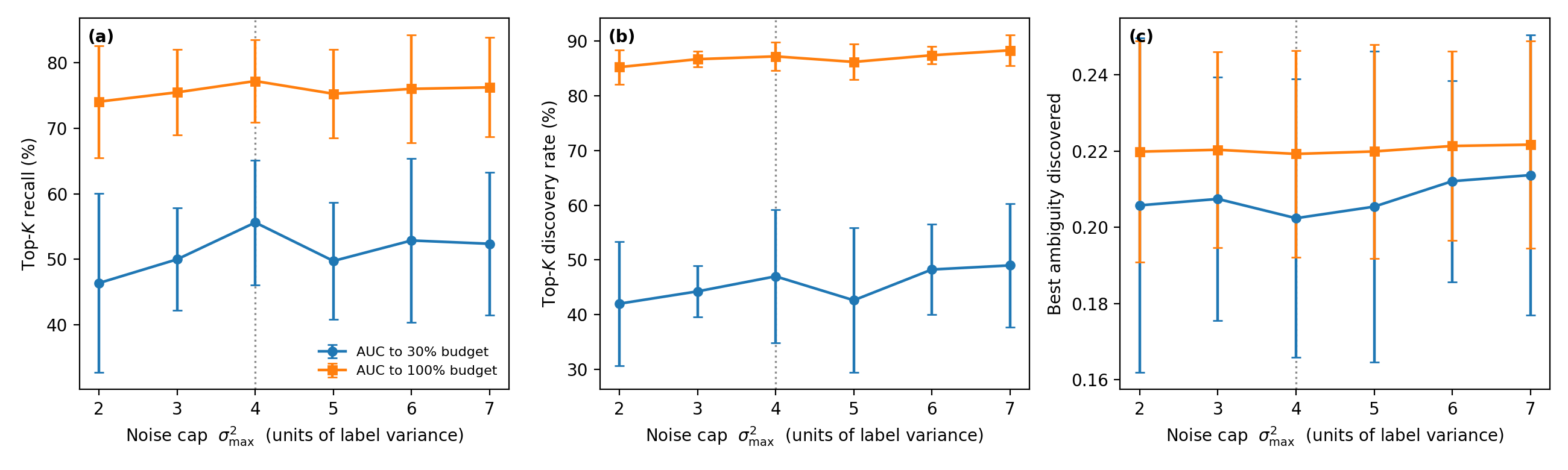}
    \caption{Sensitivity of active-learning performance to the maximum observation variance $\sigma^2_{\max}$ in the jackknife noise model, on the synthetic dataset. (a) Top-$K$ recall (b) Top-$K$ discovery rate (c) Best ambiguity discovered. Circle markers show the performance at 30\% of the pool sampled, and square markers show the performance once 100\% of the pool is sampled. Markers are means over five seeds and error bars are one standard deviation. The dotted line marks the value $\sigma^2_{\max}=4$ used in the paper. Error bars overlap across the full range, indicating that performance is not sensitive to this bound.}
    \label{fig:si_max_var}
\end{figure}

We assessed the sensitivity of the method to this bound by sweeping $\sigma^2_{\max} \in \{2,3,4,5,6,7\}$ on the synthetic dataset, with five random seeds per value (Fig.~\ref{fig:si_max_var}). All caps share the same seeds, patch pools and ground truth, so the comparison between values are fair. Across the full range, no cap differs significantly from the chosen value in the paper, as their difference is significantly smaller than the difference induced by random seeds. We therefore report that performance is insensitive to $\sigma^2_{\max}$ over this range, and that the value $4$ is not a fine-tuned quantity.

\section{Effect of Reducing the GP Input Dimensionality} \label{si:pca_dim}

Ambiguity by default is computed from the raw 384-dimensional DINOv3 embedding. We tested whether a reduced representation suffices by projecting onto the first 10, 25 and 50 principal components of a cosine PCA pre-fitted on a large micrograph dataset~\cite{aversa_first_2018}, retaining 77.8\%, 88.7\% and 94.2\% of the variance respectively. Experiments were run on the synthetic dataset using 5 different random seeds, and we report the top-$K$ recall, discovery rate, and best ambiguity discovered from 0\% to 100\% of the patch sampled.

\begin{figure}[H]
    \centering
    \includegraphics[width=1\linewidth]{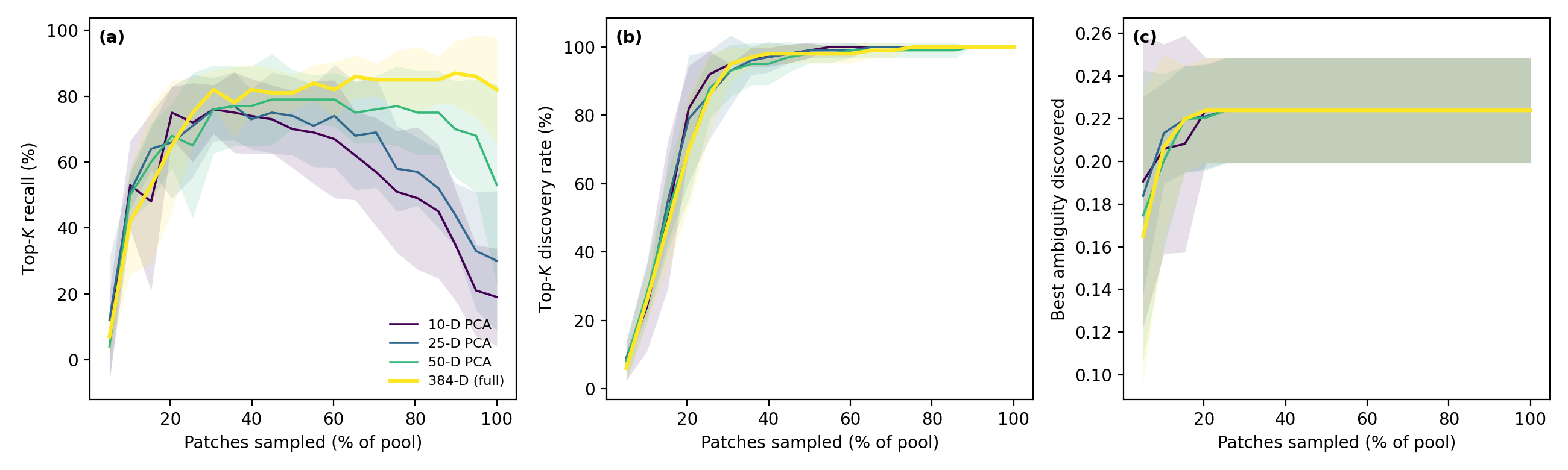}
    \caption{Active learning trajectories on the synthetic dataset for using cosine PCA reduced latent dimension of 10, 25 and 50, and the full 384-D embedding. (a) Top-$K$ recall (b) Top-$K$ discovery rate (c) Best ambiguity discovered. Lines are means over five seeds and shaded bands are one standard deviation. Discovery rate and best ambiguity discovered are unaffected by reduced latent dimensions while recall rate significantly degrades at higher percentage sampled for the reduced dimensions.}
    \label{fig:si_pca_dim}
\end{figure}

From Fig. \ref{fig:si_pca_dim}, Top-$K$ discovery rate and best ambiguity discovered is indistinguishable across the different latent dimensions at every budget, where difference is well within the random seeds variation. However, for Top-$K$ recall, the reduced, lower latent dimensions deviates from correctly predicting the highest ambiguity region as more samples are being collected. While the full 384-D version retains its ~80\% recall rate. As a pre-fitted PCA also imposes a bias on which latent dimension is important from the dataset it was fitted on, it is expected that its performance will only worsen on micrograph not seen in the dataset. Therefore, we retain the full 384-D embedding for ambiguity calculation.

\section{Different Acquisition Policy for Active Learning on Ambiguity} \label{si:al}
Seven acqusition policy was tested for active learning on ambiguityL: random, LFPS, Upper Confidence Bound (UCB), qUCB ($q=2$), Expected Improvenment (EI), qLogEI ($q=2$). We observe that all acqusition policy other than Random, LFPS, and UCB performs similarly, making them all suitable for constructing surrogate model to predict ambiguity. qLogNEI was chosen as it is a batched acquisition policy, allowing selection and imaging of multiple regions at once without sacrificing performance at predicting ambiguity.

\begin{figure}[H]
    \centering
    \includegraphics[width=1\linewidth]{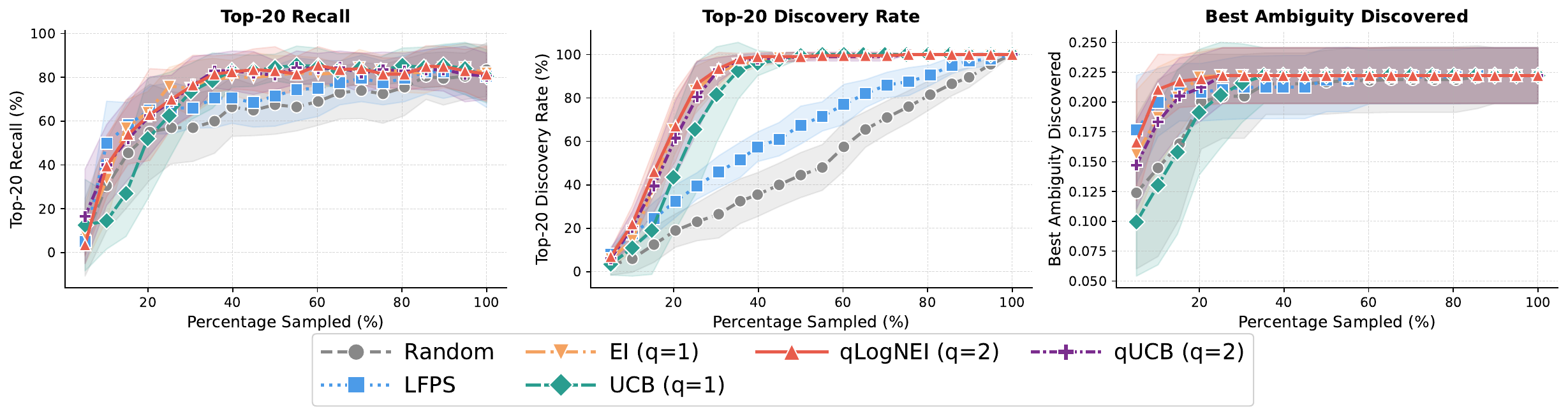}
    \caption{a)The top-20 recall, b)top-20 discovery rate, and c)best ambiguity score discovered so far, averaged over 10 different random seeds for Random, LFPS, UCB, qUCB, EI, and qLogNEI.}
    \label{fig:si_all6}
\end{figure}

\section{Performance of AutoRASOR on all 6 grid dataset} \label{si:all_grid}
\begin{figure}[H]
    \centering
    \begin{minipage}[c]{0.2\linewidth}
        \centering
        \includegraphics[width=\linewidth]{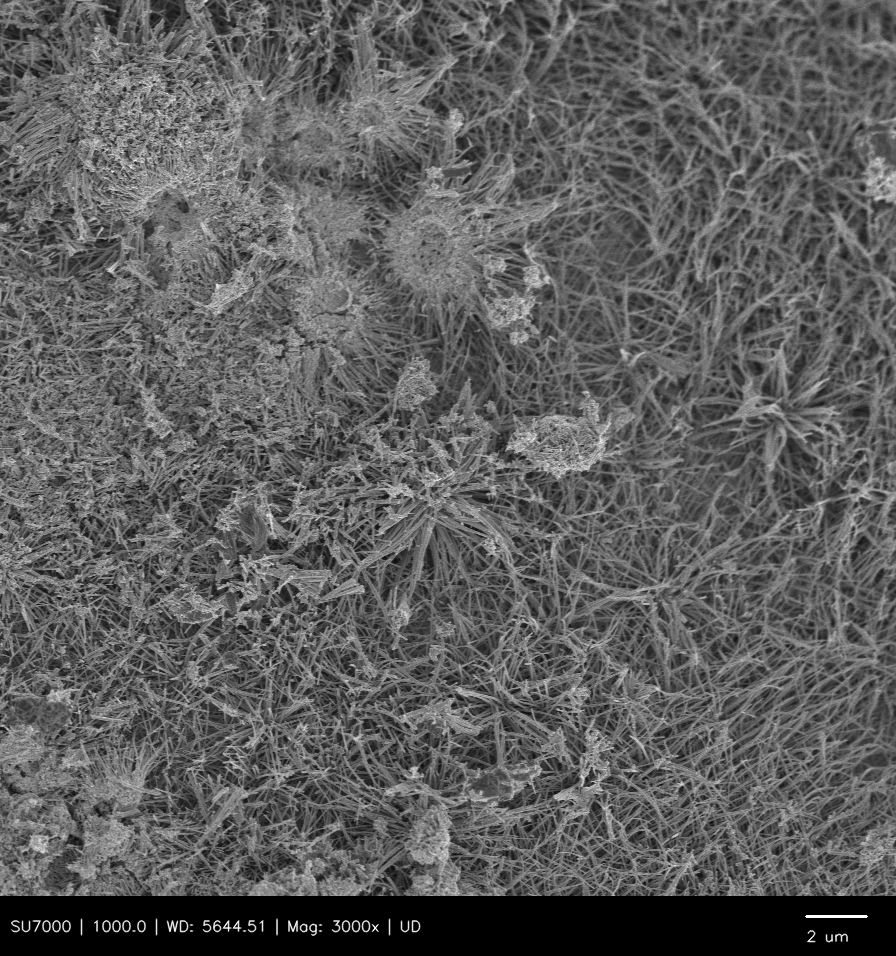}
        \label{fig:grid_7_lowmag}
    \end{minipage}%
    \hfill
    \begin{minipage}[c]{0.8\linewidth}
        \centering
        \includegraphics[width=\linewidth]{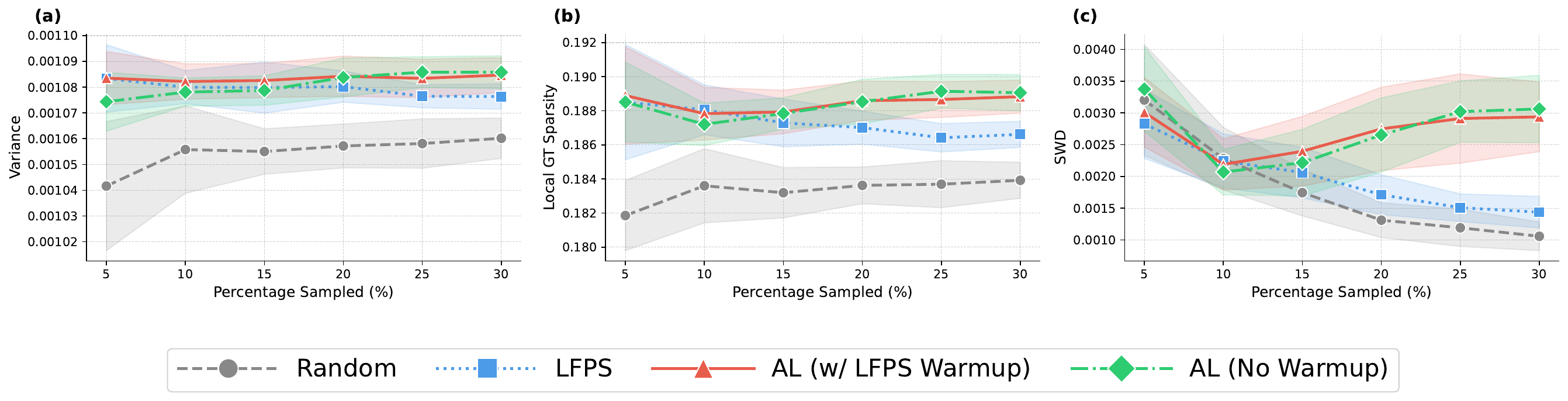}
        \label{fig:grid_7_metrics}
    \end{minipage}
    
    \caption{\textbf{Sampling policy performance for Dataset 1.} 
    \textit{Left:} Low-magnification micrograph of dataset 1. 
    \textit{Right:} Metric trajectories across acquisition budgets (5\% to 30\%) for 
    \textbf{(a)} Full-384D Latent Variance (higher is better), 
    \textbf{(b)} Local GT Sparsity (higher is better), and 
    \textbf{(c)} Sliced Wasserstein Distance (lower is better). 
    Curves represent the mean across 10 random seeds, and shaded regions indicate $\pm 1\,\text{std}$.}
    \label{fig:si_grid_7_evaluation}
\end{figure}

\begin{figure}[H]
    \centering
    \begin{minipage}[c]{0.2\linewidth}
        \centering
        \includegraphics[width=\linewidth]{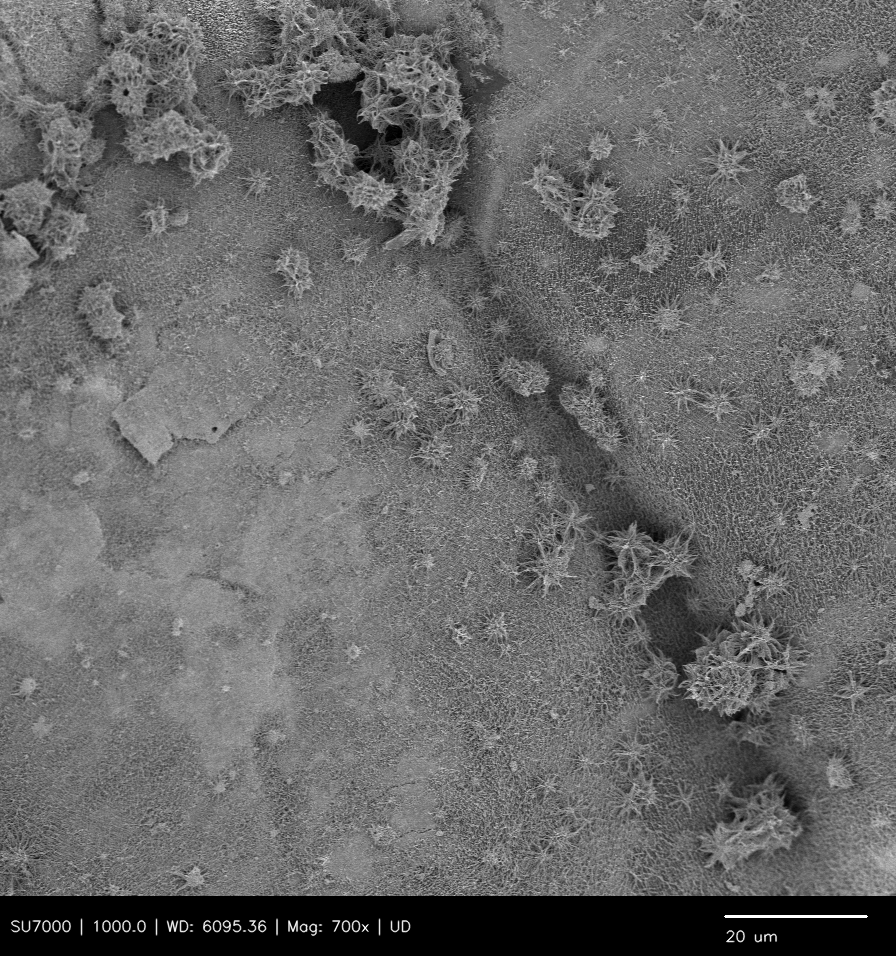}
        \label{fig:grid_8_lowmag}
    \end{minipage}%
    \hfill
    \begin{minipage}[c]{0.8\linewidth}
        \centering
        \includegraphics[width=\linewidth]{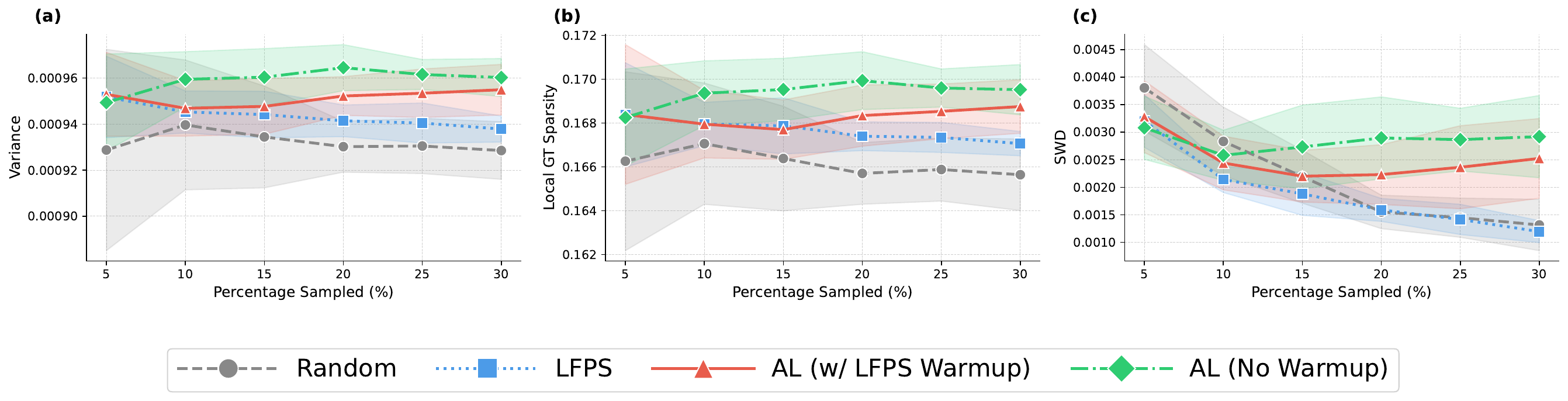}
        \label{fig:grid_8_metrics}
    \end{minipage}
    
    \caption{\textbf{Sampling policy performance for Dataset 2.} 
    \textit{Left:} Low-magnification micrograph of dataset 2. 
    \textit{Right:} Metric trajectories across acquisition budgets (5\% to 30\%) for 
    \textbf{(a)} Full-384D Latent Variance (higher is better), 
    \textbf{(b)} Local GT Sparsity (higher is better), and 
    \textbf{(c)} Sliced Wasserstein Distance (lower is better). 
    Curves represent the mean across 10 random seeds, and shaded regions indicate $\pm 1\,\text{std}$.}
    \label{fig:si_grid_8_evaluation}
\end{figure}

\begin{figure}[H]
    \centering
    \begin{minipage}[c]{0.2\linewidth}
        \centering
        \includegraphics[width=\linewidth]{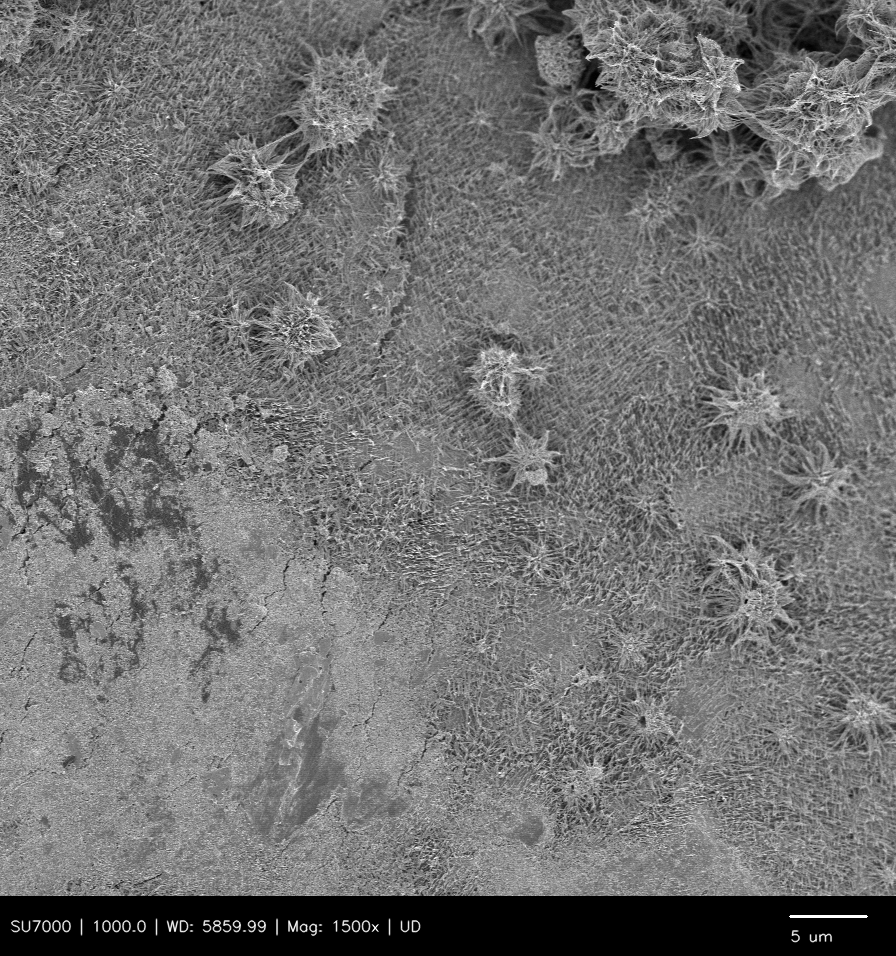}
        \label{fig:grid_9_lowmag}
    \end{minipage}%
    \hfill
    \begin{minipage}[c]{0.8\linewidth}
        \centering
        \includegraphics[width=\linewidth]{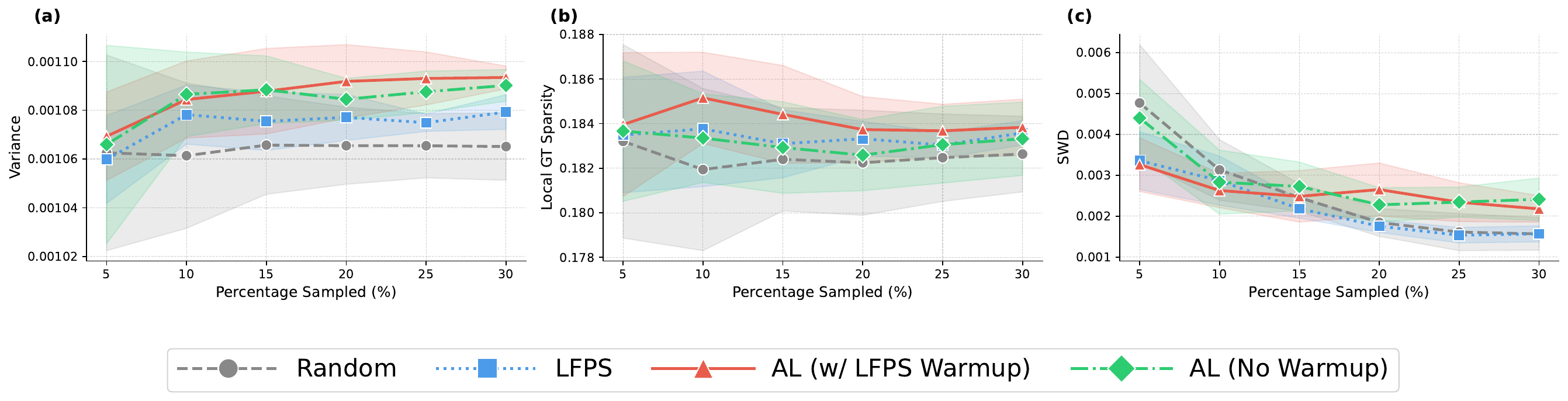}
        \label{fig:grid_9_metrics}
    \end{minipage}
    
    \caption{\textbf{Sampling policy performance for Dataset 3.} 
    \textit{Left:} Low-magnification micrograph of dataset 3. 
    \textit{Right:} Metric trajectories across acquisition budgets (5\% to 30\%) for 
    \textbf{(a)} Full-384D Latent Variance (higher is better), 
    \textbf{(b)} Local GT Sparsity (higher is better), and 
    \textbf{(c)} Sliced Wasserstein Distance (lower is better). 
    Curves represent the mean across 10 random seeds, and shaded regions indicate $\pm 1\,\text{std}$.}
    \label{fig:si_grid_9_evaluation}
\end{figure}

\begin{figure}[H]
    \centering
    \begin{minipage}[c]{0.2\linewidth}
        \centering
        \includegraphics[width=\linewidth]{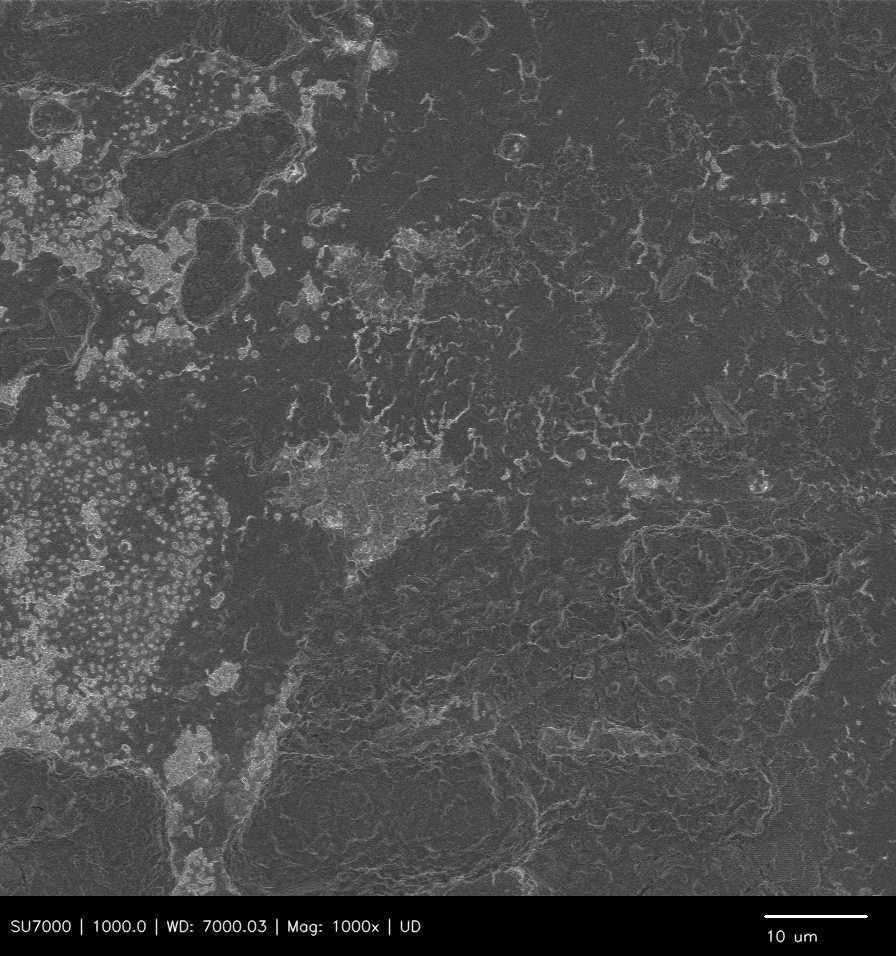}
        \label{fig:grid_10_lowmag}
    \end{minipage}%
    \hfill
    \begin{minipage}[c]{0.8\linewidth}
        \centering
        \includegraphics[width=\linewidth]{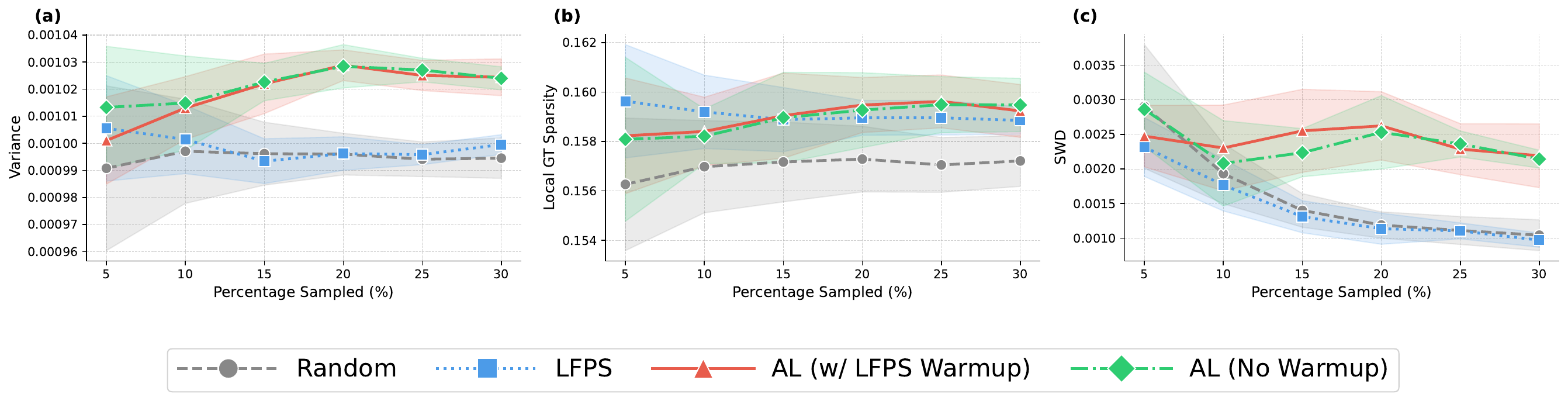}
        \label{fig:grid_10_metrics}
    \end{minipage}
    
    \caption{\textbf{Sampling policy performance for Dataset 4.} 
    \textit{Left:} Low-magnification micrograph of dataset 4. 
    \textit{Right:} Metric trajectories across acquisition budgets (5\% to 30\%) for 
    \textbf{(a)} Full-384D Latent Variance (higher is better), 
    \textbf{(b)} Local GT Sparsity (higher is better), and 
    \textbf{(c)} Sliced Wasserstein Distance (lower is better). 
    Curves represent the mean across 10 random seeds, and shaded regions indicate $\pm 1\,\text{std}$.}
    \label{fig:si_grid_10_evaluation}
\end{figure}

\begin{figure}[H]
    \centering
    \begin{minipage}[c]{0.2\linewidth}
        \centering
        \includegraphics[width=\linewidth]{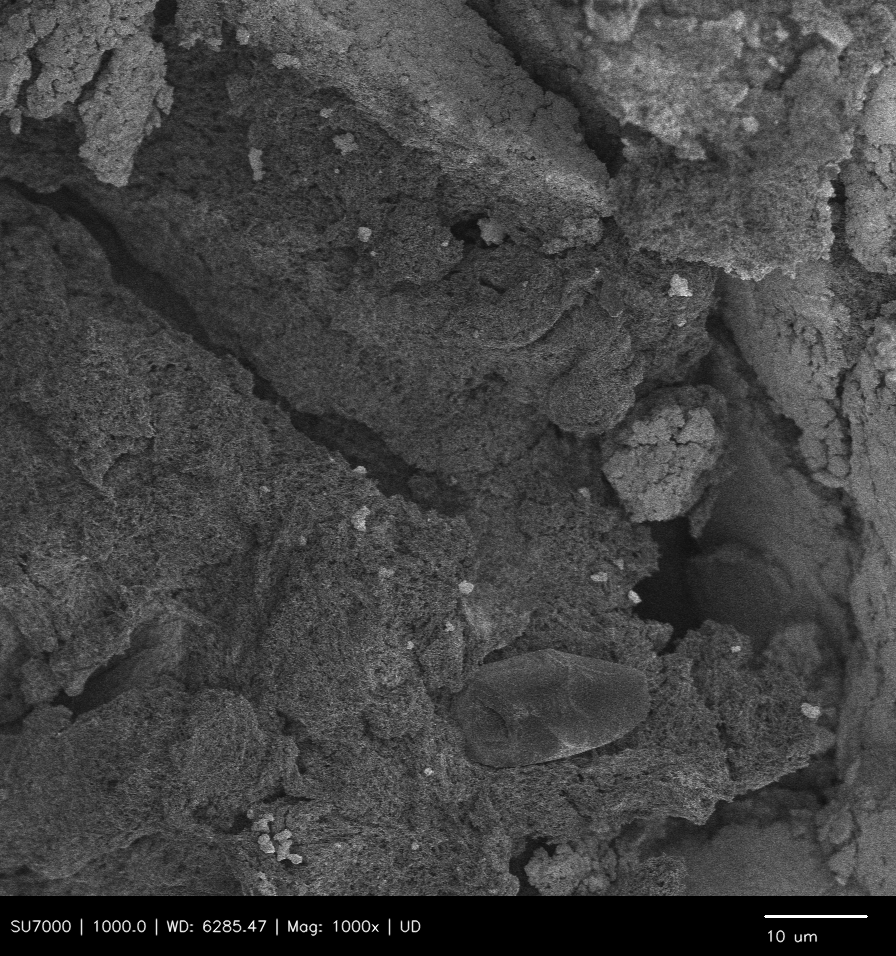}
        \label{fig:grid_11_lowmag}
    \end{minipage}%
    \hfill
    \begin{minipage}[c]{0.8\linewidth}
        \centering
        \includegraphics[width=\linewidth]{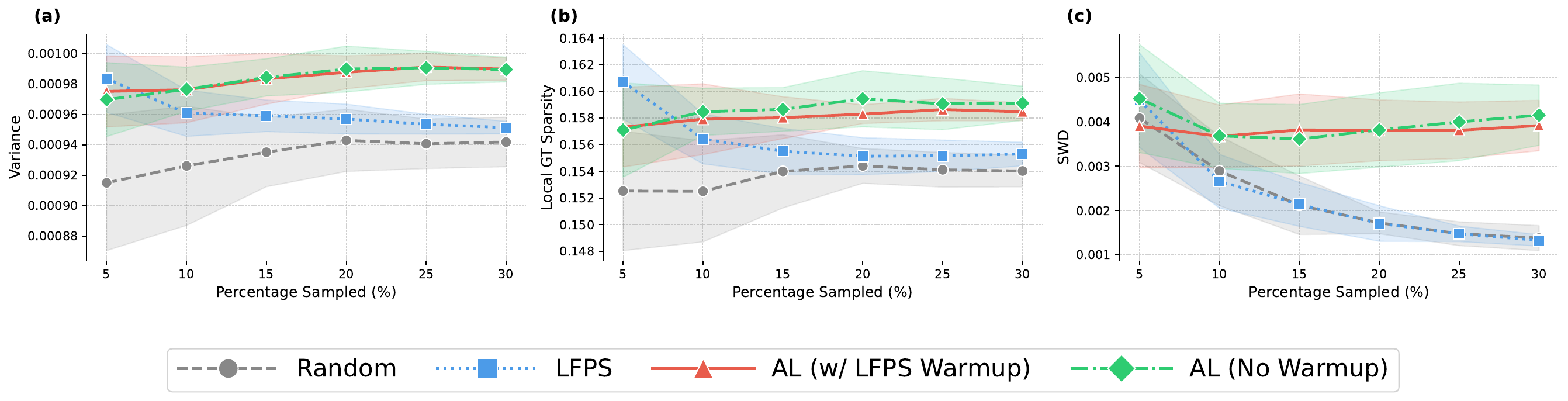}
        \label{fig:grid_11_metrics}
    \end{minipage}
    
    \caption{\textbf{Sampling policy performance for Dataset 5.} 
    \textit{Left:} Low-magnification micrograph of dataset 5. 
    \textit{Right:} Metric trajectories across acquisition budgets (5\% to 30\%) for 
    \textbf{(a)} Full-384D Latent Variance (higher is better), 
    \textbf{(b)} Local GT Sparsity (higher is better), and 
    \textbf{(c)} Sliced Wasserstein Distance (lower is better). 
    Curves represent the mean across 10 random seeds, and shaded regions indicate $\pm 1\,\text{std}$.}
    \label{fig:si_grid_11_evaluation}
\end{figure}

\begin{figure}[H]
    \centering
    \begin{minipage}[c]{0.2\linewidth}
        \centering
        \includegraphics[width=\linewidth]{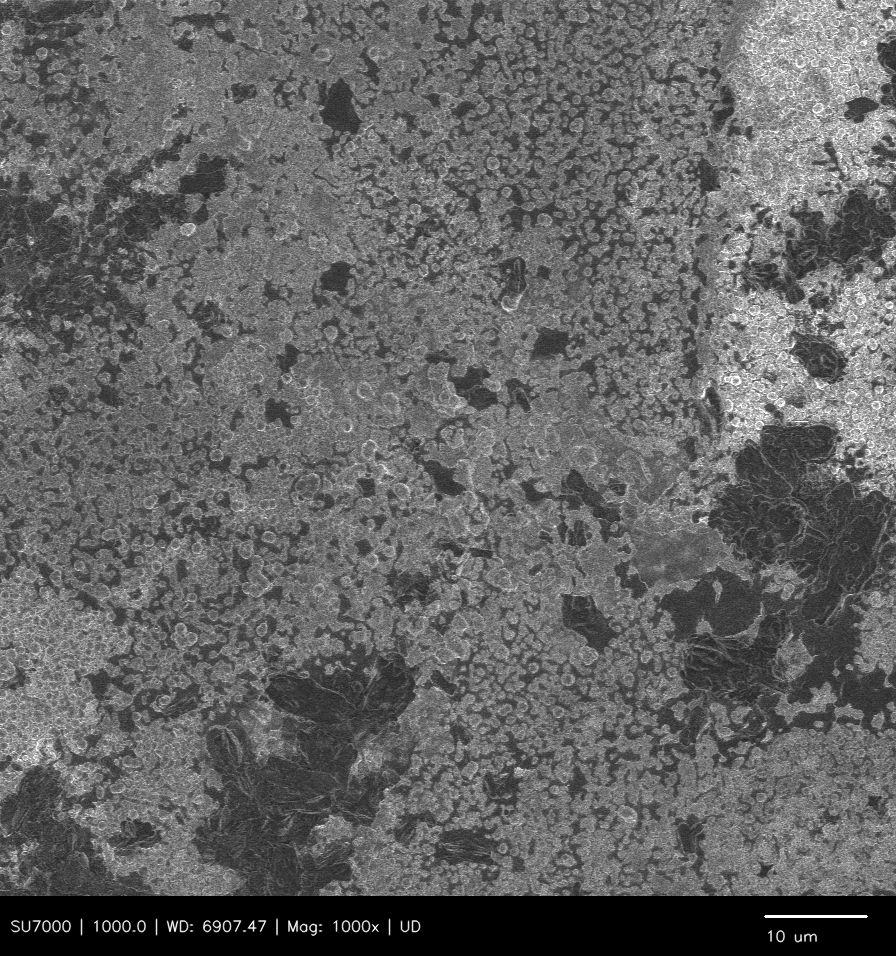}
        \label{fig:grid_12_lowmag}
    \end{minipage}%
    \hfill
    \begin{minipage}[c]{0.8\linewidth}
        \centering
        \includegraphics[width=\linewidth]{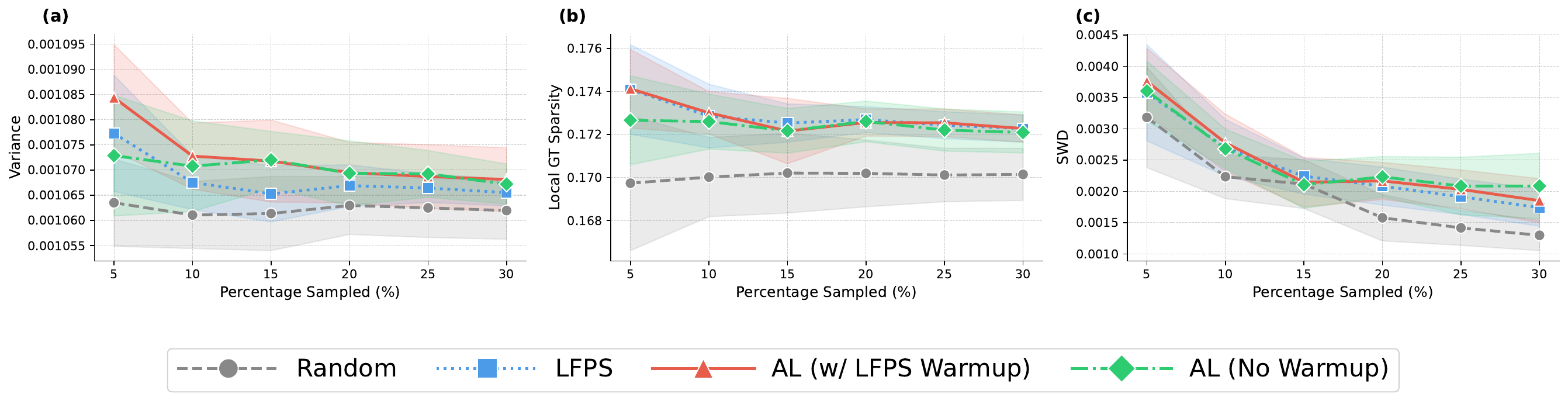}
        \label{fig:grid_12_metrics}
    \end{minipage}
    
    \caption{\textbf{Sampling policy performance for Dataset 6.} 
    \textit{Left:} Low-magnification micrograph of dataset 6. 
    \textit{Right:} Metric trajectories across acquisition budgets (5\% to 30\%) for 
    \textbf{(a)} Full-384D Latent Variance (higher is better), 
    \textbf{(b)} Local GT Sparsity (higher is better), and 
    \textbf{(c)} Sliced Wasserstein Distance (lower is better). 
    Curves represent the mean across 10 random seeds, and shaded regions indicate $\pm 1\,\text{std}$.}
    \label{fig:si_grid_12_evaluation}
\end{figure}